\documentclass[lettersize,journal]{IEEEtran}
\usepackage{amsmath,amsfonts}
\usepackage{array}
\usepackage[caption=false,font=normalsize,labelfont=sf,textfont=sf]{subfig}
\usepackage{textcomp}
\usepackage[linesnumbered,ruled,vlined]{algorithm2e}
\usepackage{algpseudocode}
\SetKwInput{kwInit}{Initialize}
\SetKwInput{kwInp}{Input}
\SetKwInput{kwOut}{Output}
\usepackage{stfloats}
\usepackage{physics}
\DeclareMathOperator{\sinc}{sinc}

\usepackage{url}
\usepackage{caption}
\usepackage{verbatim}
\usepackage{color,soul}
\usepackage{multirow}
\usepackage{graphicx}
\usepackage{cite}
\usepackage{enumitem}
\usepackage{xcolor}
\usepackage{listings}
\usepackage[flushleft]{threeparttable}
\definecolor{commentgreen}{RGB}{29,87,42}
\definecolor{eminence}{RGB}{20,26,219}
\definecolor{weborange}{RGB}{255,165,0}
\definecolor{frenchplum}{RGB}{129,20,83}

\begin{document}

\title{SQ-CARS: A Scalable Quantum Control and Readout System}

\author{%
	\IEEEauthorblockN{%
		Ujjawal~Singhal\textsuperscript{\textsection}\IEEEauthorrefmark{1},
		Shantharam~Kalipatnapu\textsuperscript{\textsection}\IEEEauthorrefmark{2},
        Pradeep~Kumar~Gautam\textsuperscript{\textsection}\IEEEauthorrefmark{2}\IEEEauthorrefmark{3},
		Sourav~Majumder\IEEEauthorrefmark{1},
		Vaibhav~Venkata~Lakshmi~Pabbisetty\IEEEauthorrefmark{2},
		Srivatsava~Jandhyala\IEEEauthorrefmark{2},
		Vibhor~Singh\IEEEauthorrefmark{1} and
		Chetan~Singh~Thakur\IEEEauthorrefmark{2} 
	}\\

 \IEEEauthorblockA{\IEEEauthorrefmark{1} Department~of~Physics,~Indian~Institute~of~Science,~Bangalore,~India-560012}\\%
 \IEEEauthorblockA{\IEEEauthorrefmark{2} Department~of~Electronic~Systems~Engineering,~Indian~Institute~of~Science,~Bangalore,~India-560012}\\
 \IEEEauthorblockA{\IEEEauthorrefmark{3} Defence~Research~and~Development~Organisation,~Bangalore,~India-560093}\\
 \IEEEauthorblockA{
 Correspondence E-mail: \IEEEauthorrefmark{1}vsingh@iisc.ac.in
 \IEEEauthorrefmark{2}csthakur@iisc.ac.in
 }
}

\maketitle 
\begingroup\renewcommand\thefootnote{\textsection}
\footnotetext{These authors contributed equally.\\}
\endgroup

\begin{abstract}

Qubits are the basic building blocks of a quantum processor which require electromagnetic pulses in giga hertz frequency range and latency in nanoseconds for control and readout. In this paper, we address three main challenges associated with room temperature electronics used for controlling and measuring superconducting qubits: scalability, direct microwave synthesis, and a unified user interface. To tackle these challenges, we have developed SQ-CARS, a system based on the ZCU111 evaluation kit. SQ-CARS is designed to be scalable, configurable, and phase synchronized, providing multi-qubit control and readout capabilities. The system offers an interactive Python framework, making it user-friendly. Scalability to a larger number of qubits is achieved by deterministic synchronization of multiple channels. The system supports direct synthesis of arbitrary vector microwave pulses using the second-Nyquist zone technique, from 4 to 9 GHz. It also features on-board  data processing like tunable low pass filters and configurable rotation blocks, enabling lock-in detection and low-latency active feedback for quantum experiments. All control and readout features are accessible through an on-board Python framework. To validate the performance of SQ-CARS, we conducted various time-domain measurements to characterize a superconducting transmon qubit. Our results were compared against traditional setups commonly used in similar experiments. With deterministic synchronisation of control and readout channels, and an open-source approach for programming, SQ-CARS paves the way for advanced experiments with superconducting qubits.

\end{abstract}

\begin{IEEEkeywords}
control-electronics, FPGA, RFSoC, superconducting qubit, quantum computing
\end{IEEEkeywords}

\section{Introduction}
\IEEEPARstart{Q}{uantum} processors utilize the properties of quantum parallelism 
and quantum interference in solving certain computational problems 
much faster than classical computers \cite{Ladd2010}.
Qubits which are the building 
blocks of quantum processors have many realizations such as 
trapped-ions \cite{zhu_trapped_2006}, semiconducting 
quantum-dots \cite{hendrickx_four-qubit_2021}, 
nitrogen-vacancy centers \cite{fuchs_excited-state_2010}, and 
superconducting qubits \cite{Devoret2013,kjaergaard_superconducting_2020} etc.
Among several such realizations, superconducting qubits are being 
aggressively pursued for scalable quantum computing platform 
\cite{arute_quantum_2019,gong_quantum_2021}.
The superconducting qubits are essentially nonlinear oscillators, 
which utilize the non-linearity of the Josephson inductance to form an 
effective quantum two-level system \cite{koch_charge-insensitive_2007}. 
These systems need microwave pulses with gigahertz (GHz) frequency 
and latency in nanoseconds for control and readout, which are 
realized by the high speed electronics at room temperature 
\cite{kjaergaard_superconducting_2020,Krantz2019,PRXQuantum.3.037001}.

The room temperature electronics that support the control and 
measurement of superconducting qubits pose three main challenges --

\begin{enumerate}
\item Direct synthesis of microwave signals: Traditionally for 
superconducting qubits, commercially available arbitrary waveform 
generators (AWG) based on Radio Frequency-Digital to 
Analog Converters (RF-DAC) with $\leq 1$ GHz of analog bandwidth 
have been used. These are usually designed for general purpose 
tests and measurements. RF qubit control pulses, which are 
typically $4-8$ GHz are obtained by upconverting the AWG 
waveforms with analog mixers. These analog mixers come with a 
Local Oscillator (LO) leakage and imperfect sideband 
suppression \cite{jolin_calibration_2020}, 
and their electrical properties vary with manufacturing tolerance and 
environmental effects. It demands a periodic calibration of 
the mixers to suppress unwanted image frequencies which is an 
overhead to the experiments \cite{jolin_calibration_2020}.

\item Scalability: The decoherence of a quantum system is generally 
attributed to fluctuations in the device and environmental 
factors. However, the decoherence of the system is not just 
internal phenomenon but also depends on the master 
clock that drives the control electronics. 
Stable and coherent signals that drive and interact 
with the quantum system reduces the error 
significantly\cite{ball2016role}. 
For a larger system, the control and readout would involve 
information processing to synthesize large number of 
control signals, estimate the state of the qubits and to 
provide a real-time feedback. In conventional systems, 
the technical details of these is often largely unavailable. 
As the number of qubits scales up, the use of these systems 
becomes challenging in terms of both cost and complexity.  
\item Lack of Unified Interface: The control and readout 
electronics come from different vendors. In order to 
operate them, the user community has to deal with 
heterogeneous user interfaces which limits the productivity.
In addition, a unified control and readout system is essential
for advance experiments involving quantum feedback or error correcting
codes to minimize the feedbak latency
\cite{kjaergaard_superconducting_2020}. 
\end{enumerate}

This invites the need for a customized engineering 
solution that meets the requirement of the high data 
rates and signal processing of  quantum computing 
community while keeping the system scalable, affordable 
and easy to work with.
The integration of Field Programmable Gate Arrays (FPGA) with Radio Frequency -Digital to Analog Converter (RF-DAC) and Radio Frequency -Analog to Digital Converter (RF-ADC) have led to several breakthroughs
such as active-reset, pulse-routing \cite{asaad_independent_2016}, 
faster readout \cite{heinsoo_rapid_2018}, stabilization
of Rabi-oscillations \cite{vijay_stabilizing_2012}, quantum error 
correction \cite{ofek_extending_2016,campagne-ibarcq_using_2016}. 
These hardware implementations have reached sufficient maturity and several commercial 
products from different vendors are also available \cite{quant,zurich,key}.  

Recently, the availability of RF-Dataconverters with high sampling 
rates in the order of gigahertz frequencies has gained interest\cite{pulipati2020xilinx,goldsmith2020control,michalak2022universal}. 
The Xilinx Zynq Ultrascale+Radio Frequency System-On-Chip 
(RFSoC) \cite{rfsoc} which is a family of devices of Field 
Programmable Gate Arrays (FPGA) comes with  Digital to 
Analog Converters (DACs) and Analog to Digital Converters (ADCs) 
of very high sampling rates. 
It also includes up/down frequency converters using internal 
digital mixers which eliminates the need for external analog mixers. 
This first generation of RFSoC device XCZU28DR comes with 
eight high precision and low power DACs and ADCs with 
maximum sampling  rates of 6.554 GSPS and 4.096 GSPS, respectively. 
These data converters are configurable and integrated with Programmable 
Logic (PL) resources of the RFSoC through AXI interfaces. 
The eight DACs  are clocked by primary onboard reference 
Phase Locked Loop (PLL) LMK04208 and onboard RF PLL LMX2594 
to generate the sample clocks \cite{ug1271} of the data converters. 
The ZCU111 evaluation board comes with a sister card XM500 
on which, two DACs and two ADCs routed to High Frequency (HF) baluns 
with –1dB Pi pad attenuators, two DACs, and two 
ADCs routed to Low Frequency (LF) baluns with -3dB Pi pad attenuators, and 
remaining four DACs, and four ADCs routed to SMAs for use 
with external custom baluns and filters, all being routed 
to Sub Miniature Version A (SMA) connectors \cite{ug1271}. The baluns are primarily added to attenuate 
higher image frequency signals generated by DAC. The RFSoC board has been utilized to demonstrate their applicability for quantum computing in recent works\cite{xu2021qubic,stefanazzi2022qick,yang2022fpga,park2022icarus,tholen2022measurement }. However, these works do not address the issues of scalability, direct synthesis of microwave pulses, and a user-friendly interface entirely.

This work utilises a single FPGA board (ZCU111 by Xilinx)
that is populated with the XCZU28DR device to develop an 
integrated framework to support a scalable Quantum Control system. 
The proposed framework, Scalable Quantum Control and Readout System (SQ-CARS) supports up to four qubits and can be 
easily scaled to control higher number of qubits. 
The main contributions of this work are - 
\begin{enumerate} 
\item Phase synchronisation of all channels using Multi-Tile Synchronisation
\item Direct digital synthesis of microwave pulses using Mix mode technique
\item Arbitrary waveform generation and lockin detection to microwave quadratures
\item A python based programming interface to configure and control the above functionalities
\end{enumerate}

The rest of the paper is organised as follows. Section II describes the system level architecture of SQ-CARS along with the design and functionalities of its individual blocks. Section III discusses the performance characterisation of SQ-CARS in which we first benchmark the performance of the 
Continuous Wave (CW) microwave signals at room temperature 
by measuring various parameters like spurious-free dynamic 
range (SFDR), single-sideband phase noise and the 
reduction in amplitudes while generating 
signals in multiple Nyquist zones, latency of RF-SoC pipeline,  multichannel phase synchronisation and comparison of the proposed system with state-of-the-art platforms. In section IV, to show the applicability of our technique, we generate the control 
pulses required for the control of the superconducting qubit
and carry out coherence measurements of a transmon qubit. Finally, conclusion is drawn in Section V.

\section{System architecture of SQ-CARS }
This section describes the overall architecture of the 
control and readout system, its submodules and their 
functionalities.  

SQ-CARS is designed in modular way with both hardware (FPGA logic) 
and firmware easily scalable for supporting more number of qubits. 
The current design which is publicly available on 
GitHub\cite{qcelex_iisc} can support up to 4 qubits. 
This limitation is imposed by the number of DACs and ADCs 
available on the board. With Channel level modularity 
SQ-CARS can be easily scaled for control and readout of more 
number of qubits either by increasing the number of DACs and 
ADCs or by frequency multiplexing the available channels.

\subsection{\centering PYNQ based User Interface}
\begin{figure*}[!htb]
\centering
\includegraphics[scale=0.8]{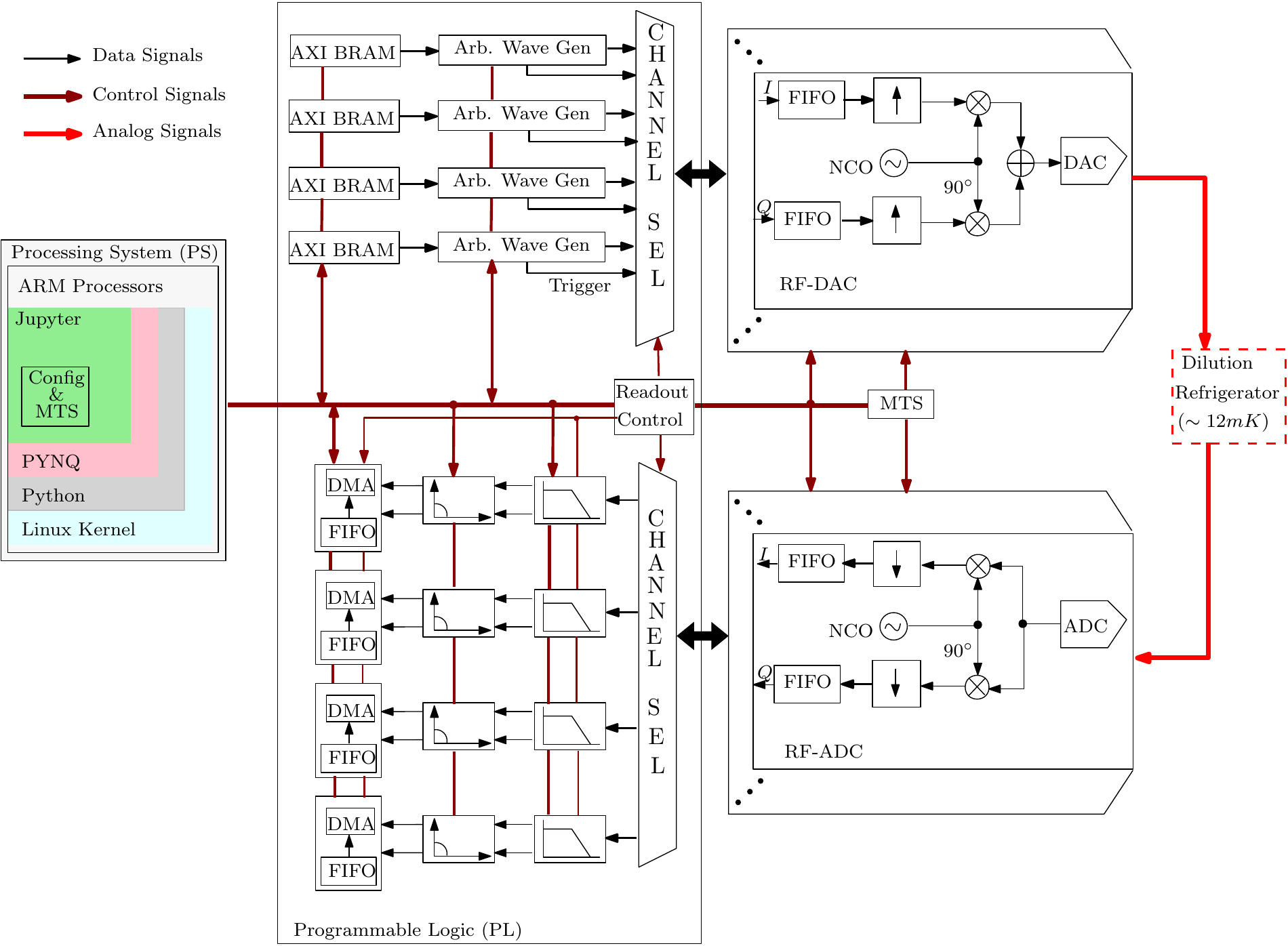}
\caption{System Architecture of $4$-Qubit SQ-CARS }
\label{Arch}
\end{figure*}

The Python based framework allows us to create an 
abstraction layer, which masks the underlying details 
of the hardware implementation and offers clean and 
user friendly configuration interface to the 
physics user community. 
The  whole architecture of the system is divided 
between Programmable Logic (PL) and the Processing System (PS) 
as shown in Figure~\ref{Arch}. ZCU111 is based on XCR28DR SoC, 
which has multicore ARM processor (PS) and Programmable 
FPGA Logic (PL). The PS runs Python productivity for 
Zynq (PYNQ) framework on Linux OS and can configure 
and control the PL design using the Overlays.

We utilized the available RFdc python code\cite{goldsmith2020control,pynqdoc} and made modifications to incorporate Multi-Tile Synchronization (MTS). Python classes were created for other hardware IPs on the PL to facilitate communication between the framework and hardware design modules. A hierarchical design approach is adopted to ensure a well-structured system  which would be beneficial as the number of qubits and their corresponding IPs scale up.

\begin{figure*}[!htb]
\centering
\includegraphics[scale=0.85]{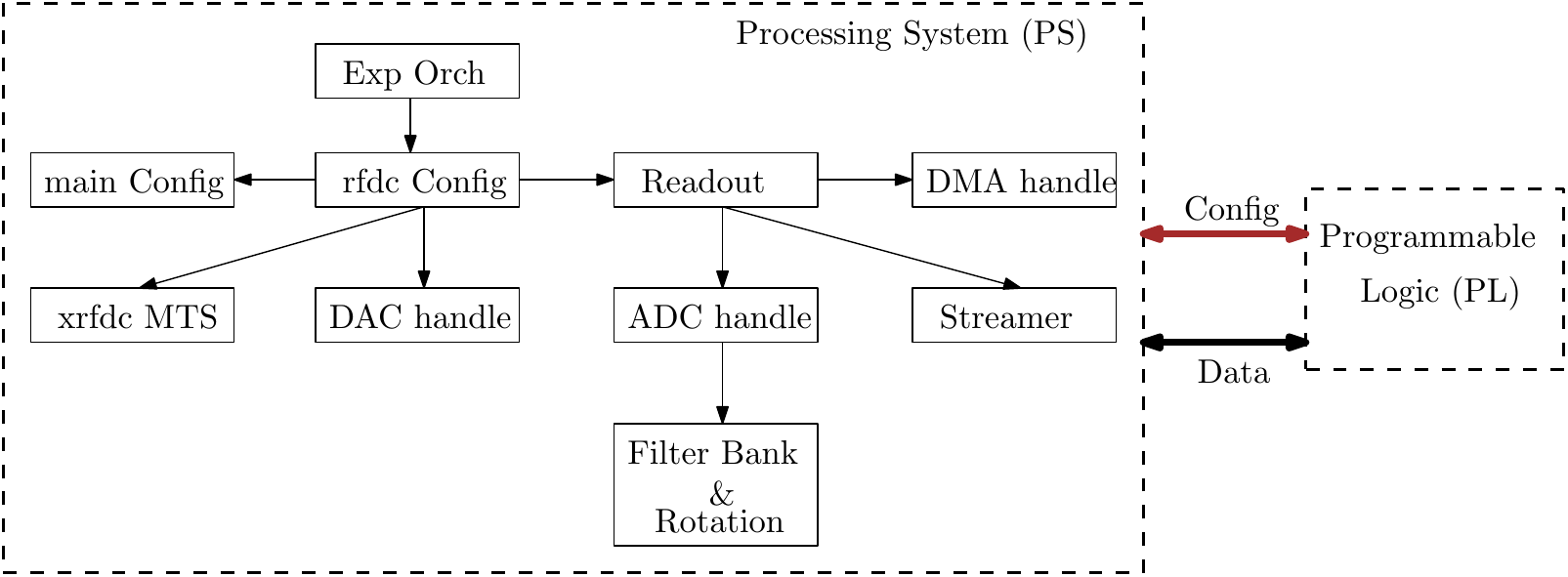}
\caption{The Class hierarchy of SQ-CARS Python Framework }
\label{PS_RFDC}
\end{figure*}

Figure~\ref{PS_RFDC} illustrates the block diagram of 
the framework running on the PS and its logical connection 
with the SQ-CARS hardware design running on the PL. 
The communication between the PS and PL can be divided 
into two parts: the Configuration \& Sync channel, and 
the data channel. Both channels are implemented using AXI 
interfaces available between the PS and PL. 
The Configuration \& Sync channel employs lightweight communication 
using Memory Mapped Input Output (MMIO) PYNQ library for read/write operations. 
The data channel utilizes the AXI Direct Memory Access (DMA) 
driver to collect data from the PL. The modular framework 
can be easily adapted for other versions of the RFSoC board 
and can be initialized and used as a Jupyter notebook or 
a Python script.
The Experiment Orchestrator (Exp Orch) shown in Figure \ref{PS_RFDC} receives the experimental 
parameters and controls the overall flow of the experiment. 
The mainConfig class holds the system configuration, such as the 
bitstream file, frequency settings, amplitude settings, 
remote host and port information, trigger settings, and more. 
The rfdcConfig class is responsible for configuring the RF data converters. 
It creates handles for the DAC and ADC channels, assigns Block 
RAM (BRAM) addresses, and sets up the RF data converter for operation. 
This class also includes methods for initializing and running 
the Multi-Tile Synchronization (MTS) process. 
The DAC and Readout classes handle the Signal Generation (AWG) 
and Readout pipeline respectively. They set NCO frequencies 
and phases, configure the DMA, and manage the data streaming 
to a remote host. The Experiment Orchestrator initiates the 
experiment by loading the waveform into the corresponding 
DAC's BRAM. Once the experiment is started, no further 
communication between the framework and PL is required, 
which reduces the experiment duty cycle.

Listing~\ref{lst:Code1} shows snippet of the code used 
for taking experiment parameter input from the user. 
Snippet shown in Listing~\ref{lst:Code2} shows the 
initialization process by providing the handle to user 
for modifying various parameters on the fly according to
the requirement of the experiment. 

\begin{lstlisting}[float,caption={Listing showing the Configuration of Experiment by the user},label={lst:Code1},captionpos=b]
#all timing parameters are in nano sec
exp_config = {
  "exp_type": "T1",
  "continuous": 1,
  "qubit_freq": 4690.2968955,
  "readout_freq": 5962.36,
  "mode": 0,
  "repetition_rate": 300000,
  "time_between_pulses": 1000,
  "initial_amp": 10,
  "trigger_delay": 0,
  "trigger_width": 4000,
  "amplitude_factor": 30,
  "amplitude_steps": 70,
  "gaussian_sigma": 400,
  "gaussian_pulse_duration": 900,
  "outer_loop_count": 1,
  "inner_loop_count": 10000,
  "inner_loop_step": 0,
  "data_fetch_time": 500000,
  "loopback": 0,
  "wave_type": "gaussian"
}

\end{lstlisting}

\begin{lstlisting}[float,caption={Listing to change the experimental parameters in Python framework},label={lst:Code2},captionpos=b]
thisConfig = SQ_CARS.mainConfig(SQ_CARS.config)
rfdc_handle = rfdcConfig(thisConfig)
# Example of setting mixer frequency of DAC
for i in rfdc_handle.dac_channels:
    # Generation of wave
    w1_I = gen_wave(thisConfig.exp_config["gaussian_pulse_duration"], thisConfig.exp_config["gaussian_sigma"])
    
    # Setting NCO Frequency
    rfdc_handle.dac[i].set_nco_freq(thisConfig.exp_config["qubit_freq"])
for i in rfdc_handle.adc_channels:    
    # Setting up DMA and streamer
    rfdc_handle.readout[i]._dma.init_dma()
    rfdc_handle.readout[i].init_streamer()
    
    # Example of passing the value to IP blocks  
    rfdc_handle.readout[i]._adc_pipeline._filter.set_iir_params(0.53) #Filter cutoff freq in MHz
    rfdc_handle.readout[i]._adc_pipeline._filter.set_iir_order(1) # Filter order
    
# Call to MTS routine and starting the experiment
rfdc_handle.run_MTS()
rfdc_handle.run_exp()

for i in rfdc_handle.adc_channels:
    jobs.new('rfdc_handle.readout[i].dma_streamer_thread()')
    rfdc_handle.readout[i].set_readout_update(1)
    jobs.status()
\end{lstlisting}

\subsection{\centering Arbitrary Waveform Generation and Control}

The control side of superconducting quantum system 
involves generation of microwave pulses of various 
shapes and duration depending on the experiment at hand. 
DACs of this board are packaged into 2 tiles 
(Tile-0 and Tile-1), each tile containing 4 DACs, 
thus providing total 8 channels. 
Tile-0 Channels are used to generate control 
microwave pulses, while Tile-1 DACs are used to 
generate corresponding readout pulses. 
Tile-1 DACs can also be used independently to 
generate control pulses, extending the design's 
capability to control upto 8-qubits. 
\par
\begin{figure*}[!htb]
\centering
\includegraphics[scale=0.85]{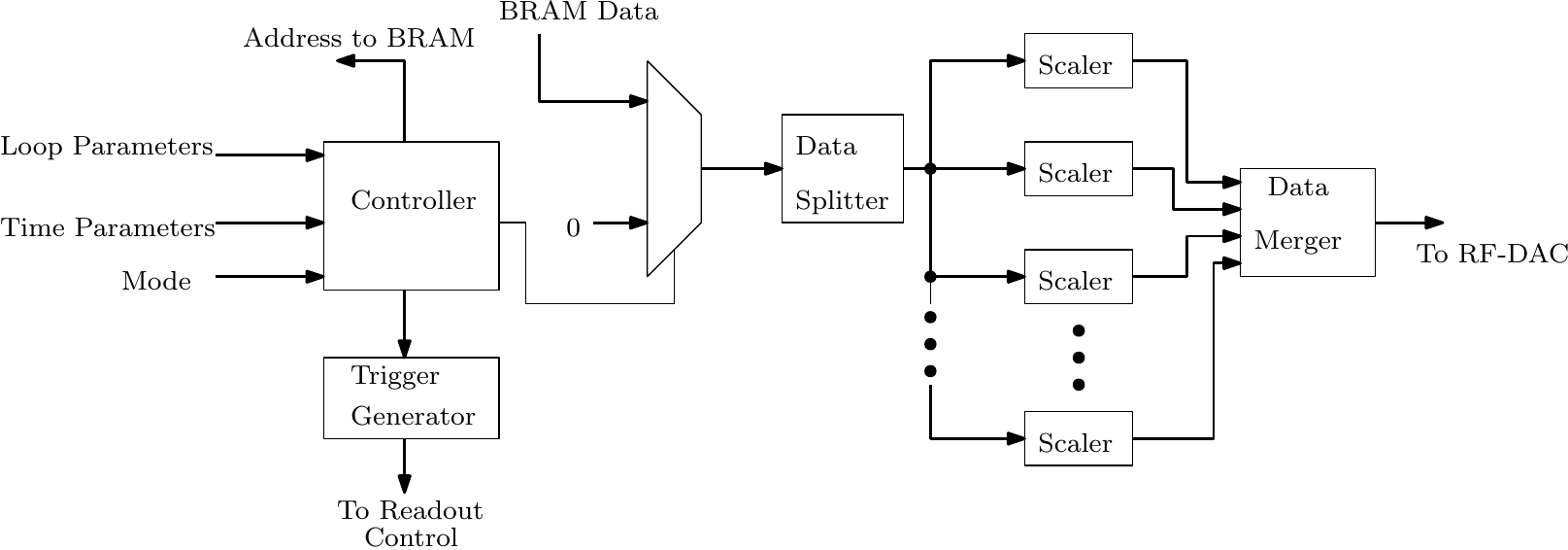}
\caption{Block Diagram of Arbitrary Waveform Generation and Control}
\label{SGC}
\end{figure*}

Each of the DACs can be controlled independently by 
the \textit{Arbitrary Wave Generator} (AWG) block which is 
shown in Figure~\ref{SGC}. 
The parameters that control the shape, amplitude, 
duration etc. of the microwave signals are controlled 
by this block. It can generate arbitrary waveform 
by reading continuous samples from local memory of 
PL (BRAM). This allows for real-time play of the 
samples and provides control to the user for changing 
the shape and duration of the waveform. These samples 
are written from Python framework along with other 
experimental parameters like mode, loop and time delays 
to the BRAM. The available local memory in PL is $4.75$MB of BRAM. 
Each sample needs two bytes of storage. 
When all the eight DAC channels are used, a maximum 
storage of $311$K samples can be allotted for each 
channel. The current design supports $64$K 
samples per channel, which is equivalent to $80~\mu$s 
of playtime at the current FPGA clock frequency of $192$~MHz, 
which is a sufficient number for most experimental purposes.  
The signal generator has the capability to play the flat 
part or time between pulses of the waveform utilizing an 
internal counter, without actually storing the respective 
portions of it in BRAM. This effectively allows us to play 
pulses of longer duration more than 80~$\mu$s in experiments 
like Time Rabi and Ramsey experiment.   
\par
The \textit{mode} allows to play continuous waveform 
or a fixed number of pulses depending on the experiment 
requirement, thus catering to a wide spectrum of quantum 
experiments. The \textit{loop} parameters help to run 
an experiment multiple times repeatedly to collect as 
many measurements as required.
The \textit{Controller} 
also controls the generation of a  trigger which 
facilitates for the timely capture on the readout 
side. The \textit{timing} parameters decide the delay of 
the trigger and width of the arbitrary wave pulses 
which are fully programmable.  The sample data read from BRAM is sent to 
\textit{Data Splitter} and the amplitude of each 
sample is scaled by the \textit{Scaler} block. 
These scaled samples are further merged and the 
interleaved $I$ and $Q$ samples are fed to the RF-DAC. 
The \textit{Channel Select} allows us to select whether 
control or readout pulses be played through DACs.
 
The \emph{RF-DAC} consists 
of a First In First Out (FIFO), an interpolation filter, a mixer, 
and a DAC as shown in Figure \ref{Arch}. 
The scaled sample values from the \emph{Arbitrary Wave Generator} are 
fed into AXI Stream (AXIS) FIFO of the corresponding DAC channel. 
At higher sampling rates of Data Converters, the data 
streaming clock of the DAC cannot be pushed in the orders 
of their sampling rates.
So a digital upsampling is necessary. This function on the 
DAC side is realized by the Interpolation Filters. The interpolation rate 
can be chosen among 1x, 2x, 4x, and 8x. 

Algorithm~\ref{ControlAlgo} shows the pseudo code of 
the control and readout that suits the needs of the 
user in characterizing a superconducting qubit. 
This algorithm is implemented as the entirety of the hardware architecture. 

Using this algorithm, basic characterizing experiments such
as the measurement of energy relaxation time, dephasing, Rabi-oscillations 
in time-domain or power domain \textit{etc.} 
can be configured easily. It takes arguments such as Ramsey, 
Time-Rabi, T$_1$, Power-Rabi, $\epsilon_{A}$, $\epsilon_{T}$, 
$\epsilon_{t}$, $Experiments$ and  $N_{iter}$ as the inputs. 
$N_{iter}$ indicates the the number of RF pulses that 
needs to be sent to the dilution refrigerator. 
The parameters $\epsilon_{A}$, $\epsilon_{T}$, and $\epsilon_{t}$ 
represent the increments in power, trigger delay and time 
between the pulses respectively. 
Depending on the kind of experiment that needs to be performed, 
these parameters are initialised and updated accordingly 
as shown in Algorithm \ref{ControlAlgo}.

\begin{algorithm}[!htb]

\caption{Pseudo code of Control and Readout}
\label{ControlAlgo}
\SetAlgoLined

\kwInp{$Ramsey$,~$Time\_ Rabi$,~$T_1$,~$Power-Rabi$, $\epsilon_{A}$,$\epsilon_{T}$,$\epsilon_{t}$, $Experiments$, $N_{iter}$}

\kwInit{$n=0$;~$N=0$;\\
        \uIf{$Power\_ Rabi$}{
        $\Delta A$ = $\epsilon_{A}$;
        $\Delta T$ = $0$;
        $\Delta t$ = $0$;
        }
\uElseIf{$T1$}{
$\Delta A$ = $0$;
$\Delta T$ = $\epsilon_{T}$;
$\Delta t$ = $0$;
}
\uElseIf{$\left(Ramsey\right)\lor\left(Time\_ Rabi\right)$}{
$\Delta A$ = $0$;
$\Delta T$ = $0$;
$\Delta t$ = $\epsilon_{t}$;
}
\Else{
$\Delta A$ = $\Delta T$ = $\Delta t$ = $0$;
}

}

\While{$\left(N<Experiments\right)$}{
        \While{$\left(n<N_{iter}\right)$}{
                Play Control Pulses;\\
                Capture Readout Data on Trigger;\\
                Process the Captured Data;\\ 
                $n=n+1$;\\
        
        }
        $A=A+\Delta A$;\\
        $T=T+\Delta T$;\\
        $t=t+\Delta t$;\\
        $N=N+1$;\\
}

\end{algorithm}

\subsection{\centering Multi-Nyquist zone operation}

According to Nyquist criteria, the sampling rate limits the frequency 
that can be faithfully reconstructed to be less than half the 
sampling rate. However, in practice, when a signal is sampled, its 
images appear at higher frequencies. Each Band of the spectrum with 
width $\frac{F_S}{2}$ is termed as Nyquist zone (NZ). For example, 
the range from DC to $\frac{F_S}{2}$ is termed as first-Nyquist zone, 
$\frac{F_S}{2}$ to $F_S$ is called second-Nyquist zone and so on \cite{ROUPHAEL2009199}. 
The images above the first-Nyquist zone can be utilized according
to the need. The output voltage of DAC can be 
represented \cite{bakercmos} as,

\begin{equation}
\label{eqn:1}
v\left(t\right) = \left[x\left(t\right)\sum_{k=-\infty}^{\infty} \delta \left(t-kT\right)\right]\ast r\left(t\right),
\end{equation}

where $x\left(t\right)$ is the desired waveform whose 
samples are being generated, $r(t)$ is the reconstruction 
waveform, and $T=\frac{1}{F_S}$. Taking Fourier transform 
on both sides of the above equation, we get

\begin{equation}
\label{eqn:2}
V\left(\omega\right)=\left[X\left(\omega\right)\ast \sum_{n=-\infty}^{\infty}\delta\left(\omega T-2\pi n\right)\right]R\left(\omega\right),
\end{equation}

where $V\left(\omega\right)$, $X\left(\omega\right)$ and $R\left(\omega\right)$ 
represent the Fourier transform of $v\left(t\right)$, $x\left(t\right)$ 
and $r\left(t\right)$, respectively.
From equation~\ref{eqn:1} and \ref{eqn:2}, it becomes evident that 
the sampled signal is passed through a system having transfer function 
$R\left(\omega\right)$ and when a signal is sampled we get its 
copies in higher frequency ranges. Therefore, the output signal strength 
of the copies in different range of frequencies gets affected 
according to the response of $R\left(\omega\right)$.

The RFSoC supports two modes of operation: Normal mode 
or Non-Return-to-Zero (NRZ) mode and Mix mode or Return-to-Complement (RTC) 
mode which determines $R\left(\omega\right)$. In the NRZ mode of operation the 
DAC uses a fixed level reconstruction waveform during one clock
cycle. It has high output power in the first Nyquist zone, but 
low output power in the second Nyquist and beyond.
RTC mode or the mixed mode, it outputs the sample for 
the first half of the clock period and then inverts the 
sample for the second half of the clock period. 
The resulting frequency response shows high power in the
second Nyquist zone and attenuation in the first Nyquist zone. 
The output power goes to zero at DC, and 2$F_S$. 
This mode provides highest power for the second Nyquist zone 
applications. Mathematically, the reconstruction waveforms 
can be written as,

\begin{equation}
	r\left(t\right)=\begin{cases}
		u\left(t\right)-u\left(t-T\right), &\text{NRZ mode}\\
		u\left(t\right)-2u\left(t-T/2\right)+u\left(t-T\right), & \text{RTC mode}
	\end{cases}
	\label{}
\end{equation}

\begin{equation}
	R\left(\omega\right)=\begin{cases}
		T e^{-i\omega T/2} \sinc \left(\frac{\omega T}{2}\right), & \text{NRZ mode}\\
		Tie^{-i\omega T/2}\sinc \left(\frac{\omega T}{4}\right)\sin \left(\frac{\omega T}{4}\right), &\text{RTC mode}
	\end{cases}
	\label{}
\end{equation}

\begin{figure*}
	\centering
	\includegraphics[scale=0.85]{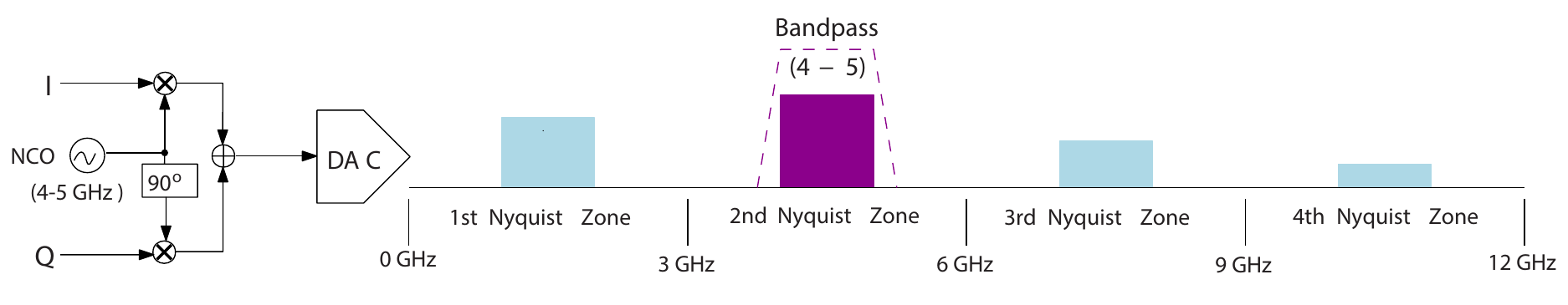}
	\caption{Up conversion scheme on RFSoC using the on-chip 
		IQ mixer and a numerically controlled oscillator. 
		Using the maximum sampling rate $F_S$~=~6.554~GHz, and 
		mix-mode of operation of DAC, the signal/image can 
		be pushed in the second Nyquist zone which can be conditioned
		using a Mini-circuit VBFZ-3590-S+(15542) band-pass filter.}
	\label{Bandfig}
\end{figure*}

The maximum sampling rate of DACs in RFSoC is $6.554$ GSPS. 
Since the frequencies necessary for the control side of qubit are 
typically in the range of $4-8$~GHz, the desired  RF band for signal 
synthesis fall in the second Nyquist zone. It can be accessed by 
operating in mix mode while maintaining highest power. 
The major advantage of this approach is that RF signal 
is generated using onboard NCOs and a digital IQ-mixer. Therefore, 
it allows a full vector control of amplitude, frequency and phase on the fly.
To suppress the images in other Nyquist zones and signal 
conditioning, we use standard coaxial band-pass filters.
In principle, these filters can be incorporated 
on a custom daughter board. A schematic of the flow and different 
Nyquist zones are shown in Figure~\ref{Bandfig}.

\subsection{\centering Readout}
On the readout side, the signals from the dilution 
refrigerator are fed to RF-ADCs. The board has four tiles, 
each tile containing two ADCs. Among the eight ADCs in total, 
four ADCs are differential ended and four ADCs are single ended. The design supports four independent readout lines. 
The user has the flexibility to chose between a single-ended 
or differential ended ADC for each readout channel which is facilitated by \textit{Channel Select}. It also allows us to lock ADC to any particular DAC channel for capturing the readout based on the internal trigger.
The ADCs are designed to directly measure the signals 
in higher Nyquist zones. The zone of operation is indicated 
to the digital calibration engine of RFSoC to ensure 
optimal performance of the ADC. 
The input samples are provided by the parallel digital 
interface of the high-speed ADC. The incoming signal is 
down-mixed using the internal NCO of RF-ADC.  
Upon down-mixing and decimation, the signal is low-pass 
filtered (LPF) and smoothed using a moving-average (MA) filter.  
The NCO frequency, cut-off frequency of the LPF can be 
configured using the Python framework based on experiment. 
The in-phase $(I)$ and quadrature $(Q)$ samples received 
after the filter are fed to Rotation block which performs,
\begin{equation}
    \begin{pmatrix}
          I^{'}\\
          Q^{'}
    \end{pmatrix} = \begin{pmatrix}
          \cos\theta & \sin\theta \\
          -\sin\theta & \cos\theta 
          
    \end{pmatrix}\begin{pmatrix}
          I\\
          Q
    \end{pmatrix} 
\end{equation}
Such a rotation block, equivalent to the auto-alignment of
phase in a lock-in measurement, is a helpful feature while performing
qubit readout using single quadrature.
The  angle of rotation can be changed by the user using 
the Python framework. This estimation can be used to determine
the qubit state or to provide an active feedback to 
the control side to further fine-tune the superconducting system.

The filter and rotation blocks can also be bypassed to 
capture the raw data. The resultant samples are fed 
into AXI Stream FIFO for transfer to PS using AXI-DMA block. 
The AXI Stream handshake signals like \textit{tvalid} 
and \textit{tlast} are generated using Readout Control Block, 
depending on the internal trigger received from Signal Generation Block. 
Stream to Memory Mapped (S2MM) port of AXI-DMA engine is 
connected to the PS via AXI Slave interface on ZYNQ SoC. 
The DMA engine is configured in Scatter Gather (SG) mode to 
facilitate use of cyclic buffer and eliminate need for PS to 
continuously provide descriptors to the DMA engine.
To fetch the descriptors, a separate SG port of DMA is used.
A custom wrapper is created in PYNQ to support DMA in 
Scatter Gather mode. Both descriptors and data buffers 
for DMA are kept in DDR memory attached to PS. 
Each readout channel is equipped with its own separate 
DMA channel, providing fine grained control on data capture 
on readout side.  The readout data received on PS is transferred 
to remote PC using the PS GEM-3 Ethernet which is either 
saved in file or played using python scripts for initial 
startup of the experiment. 
 
\subsection{\centering Multi-Tile Synchronisation}
To control and measure multiple qubits, the output 
waveforms of multiple ADC and DAC channels requires to be 
synchronised both in time and phase. This can be achieved 
by an external 10~MHz clock which is fed to LMK04208 to 
synchronize the timebase of the FPGA board with other 
instruments and Multi Tile Synchronisation (MTS).
The RF-ADC and RF-DAC constitute dual clock FIFOs. 
The data converters in a single tile share the clocking 
and data infrastructure, therefore the FIFO latency within a tile 
remains the same. However, the latency of the FIFOs can 
vary from one tile to another. MTS enables to achieve 
the relative multi-tile alignment. A single master clock 
generates all clock signals required for RF data converters 
and the programmable logic.
A custom python wrapper for C-drivers is developed 
to configure and invoke MTS.  

\section{Performance Characterisation of SQ-CARS}
This section describes various measurements and characterisation 
done with 4-qubit SQ-CARS design on ZCU111 to benchmark 
the performance of the proposed system with the state-of-the-art 
traditional setup. For all the measurements discussed in this section, the DACs and ADCs are operated at sampling rates of $6.144$~GSPS and $3.840$~GSPS respectively with their NCOs operating in Mix/Normal mode as per the requirement.

\subsection{\centering Magnitude Response of DAC}

To measure the power dependency in NRZ and RTC modes, the DAC 
output response is being characterized by enabling 
the NCO  whose frequency is changed on the fly  
and the corresponding magnitude is recorded. 
To compensate for the loss in magnitude, an inverse sinc filter 
is applied in both modes of operation. The recorded DAC output response 
is shown in Figure~\ref{RFPerformance Fig}(a). 
The DAC output response in NRZ mode follows a sinc function.
It is evident from the Figure~\ref{RFPerformance Fig} that the signal power 
is maximum in the second and third Nyquist zone when operating 
in the RTC mode. 

\subsection{\centering Multi Channel Phase Synchronisation}

The phase synchronisation between the control and readout 
channels is critical in a multi-qubit system.  
QICK\cite{stefanazzi2022qick} has reported achieving inter-channel 
phase synchronisation using the tprocessor. 
However, their approach does not consider the variable 
delay introduced by the RF-SoC pipeline, due to NCOs and FIFOs. 
To address this limitation, our work SQ-CARS, utilises MTS with on-board NCOs enabled, 
to generate and capture phase synchronised signals, demonstrating 
inter-channel synchronisation from generation by DACs to capture on ADCs.

\begin{figure}[!htb]
\begin{center}
\includegraphics[scale=0.4]{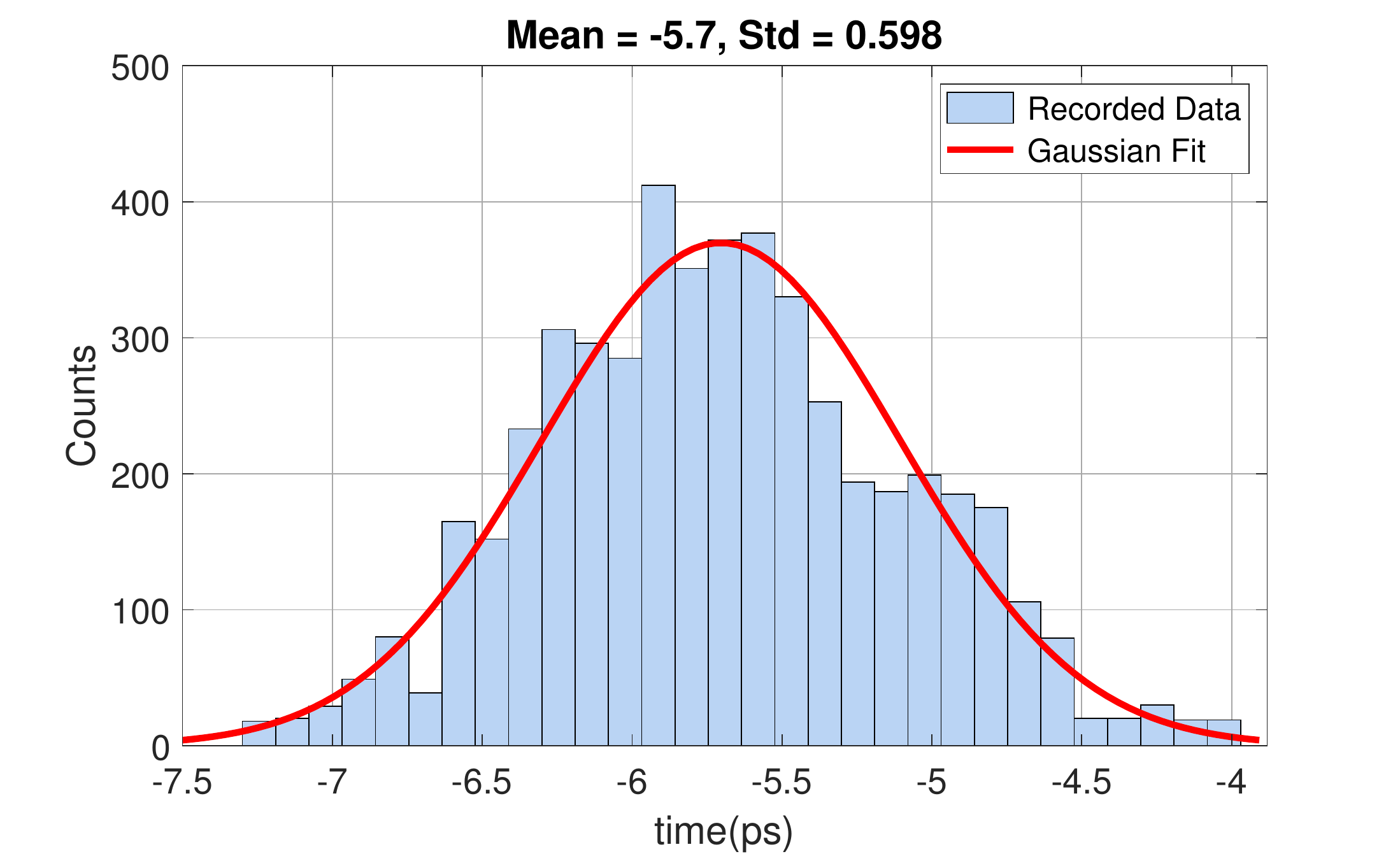}
\caption{Histogram of channel to channel jitter of two DACs}
\label{MTSJitter}
\end{center}
\end{figure}

To test the multi channel phase synchronisation on control side, 
we generate a $1$~MHz sinusoid by two DACs on ZCU111. 
These signals are being observed on an oscilloscope for 
jitter measurement. The channel to channel jitter of two 
DACs is measured by repeatedly sampling the jitter $4000$ 
times with a time interval of  $100~\mu s$ . 
The histogram of jitter is shown in Figure 
\ref{MTSJitter}. The standard deviation of channel 
to channel jitter is $\sim 0.6~ps$ which is better than $\sim 5~ps$ and   $< 1~ps$ as reported in Ref~\cite{yang2022fpga} and Ref~\cite{tholen2022measurement}, respectively.  Low standard deviation of jitter  shows stable  channel synchronisation to support high-precision synchronized qubit operations. 

\begin{figure}[!htb]
\begin{center}
\includegraphics[height=8cm, width=9cm]{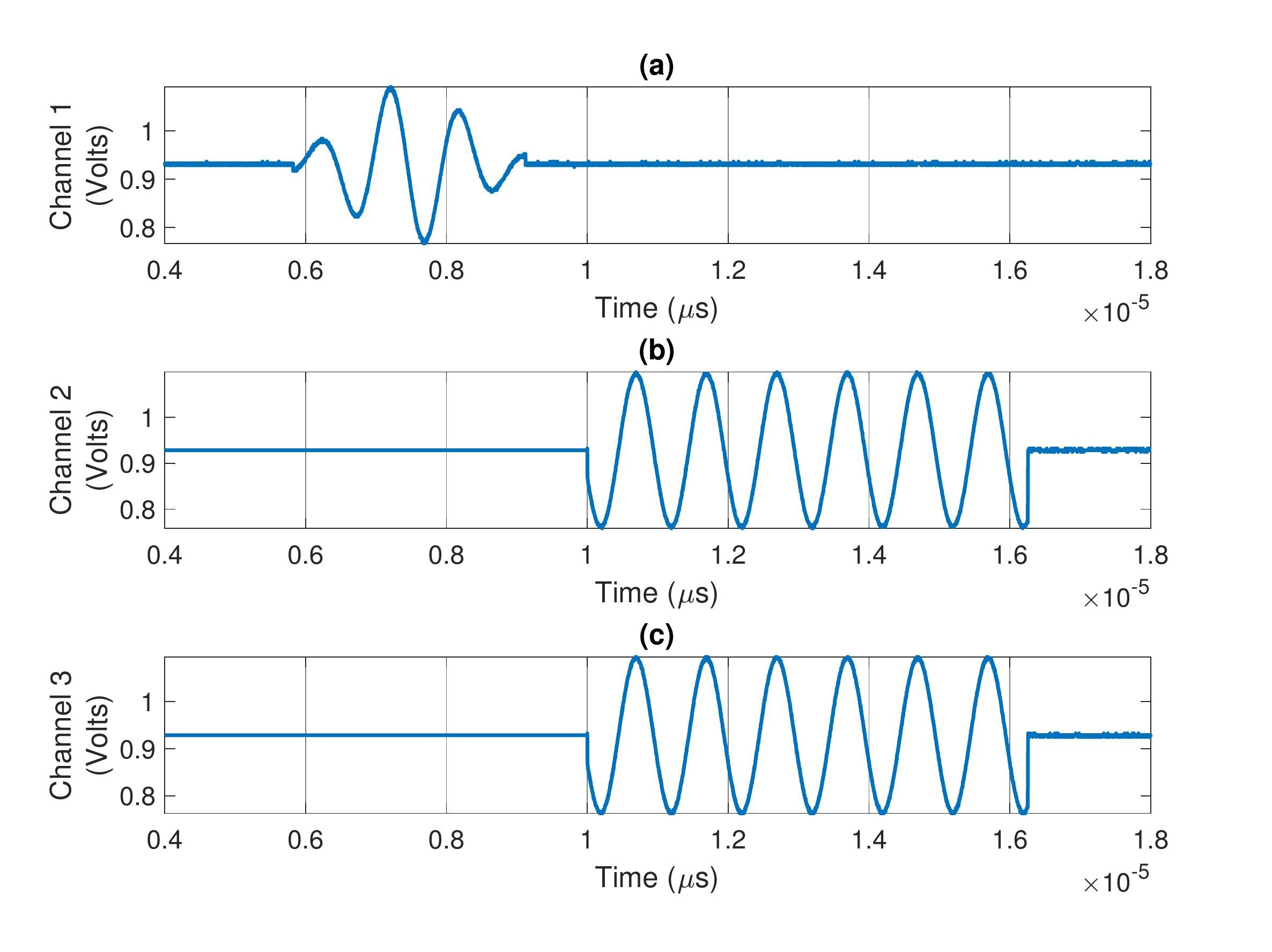}
\caption{Phase Synchronisation across various DAC Channels (a) Gaussian control pulse on Channel $1$ (b) Readout pulse on Channel $2$ (c) Readout pulse on Channel $3$. The readout pulse starts with a delay after the control pulse. It can be seen that the readout pulses on Channels $2-3$ are in synchronisation.}
\label{MTSDAC}
\end{center}
\end{figure}

\begin{figure}[!htb]
\begin{center}
\includegraphics[scale=0.3]{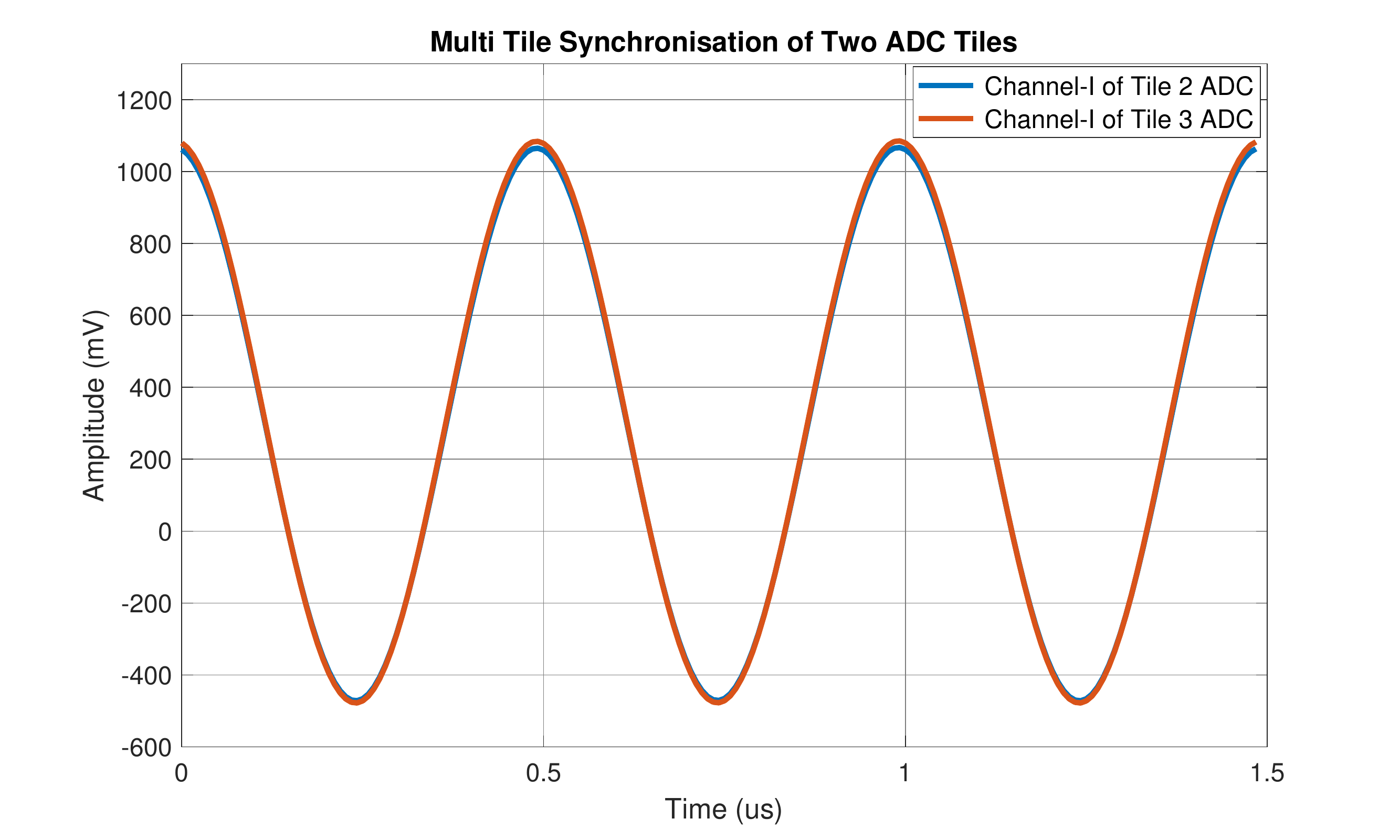}
\caption{Multi-Channel Phase Synchronisation of two ADC channels}
\label{MTSADC}
\end{center}
\end{figure}

In order to demonstrate the multi channel phase synchronised 
capture on readout side, a $1$~MHz sinusoid generated by DACs
as earlier (which are already phase synchronised) is fed to ADCs. 
The resultant signal at the output of the ADC, the $I$ and $Q$ 
channels are captured. and found to be in phase synchronisation. 
Figure~\ref{MTSADC} shows the output of  $I$-channels of two such ADCs.

\subsection{\centering Latency}

Latency is an important metric for qubit readout, active
feedback, and error-correction protocols.
The RFSoC ADC and DAC pipelines have a series of modules 
of FIFO, Interpolation/Decimation filters, Mixers as shown 
in Figure~\ref{Arch}. These modules can be used or bypassed 
as per the requirement. The latency  changes as we bypass 
or utilise these modules.  Latency measurement was done 
using loopback between DAC and ADC and observing the markers 
in Integrated Logic analyzer (ILA) running at $192$~MHz clock.  
The interpolation rate of DAC is 8x and decimation rate of ADC 
is set to 4x, the mixer is enabled and the DAC and ADC channels are 
synchronised using MTS. The round-trip latency including DAC, 
ADC and the digital AXI interfaces associated with the data 
converters is measured to be 48 cycles equalling $250$~ns with 
the design operating at $192$~MHz. 
This number is comparable to the latency measurement reported 
without MTS by Quantum Instrumentation Control Kit 
(QICK) \cite{stefanazzi2022qick}.
Each half of the measured latency is contributed by RF-DAC 
and RF-ADC pipelines.

\subsection{\centering Noise Characterisation}
Another important benchmarking metric for continuous mode 
of operation is the single-sideband phase noise. A frequency tone 
(carrier) is generated, and at various offset frequencies 
from the carrier, the power is measured in a specified bandwidth 
of 1~Hz.

\begin{figure}[!htb]
\centering
\includegraphics[scale=0.52]{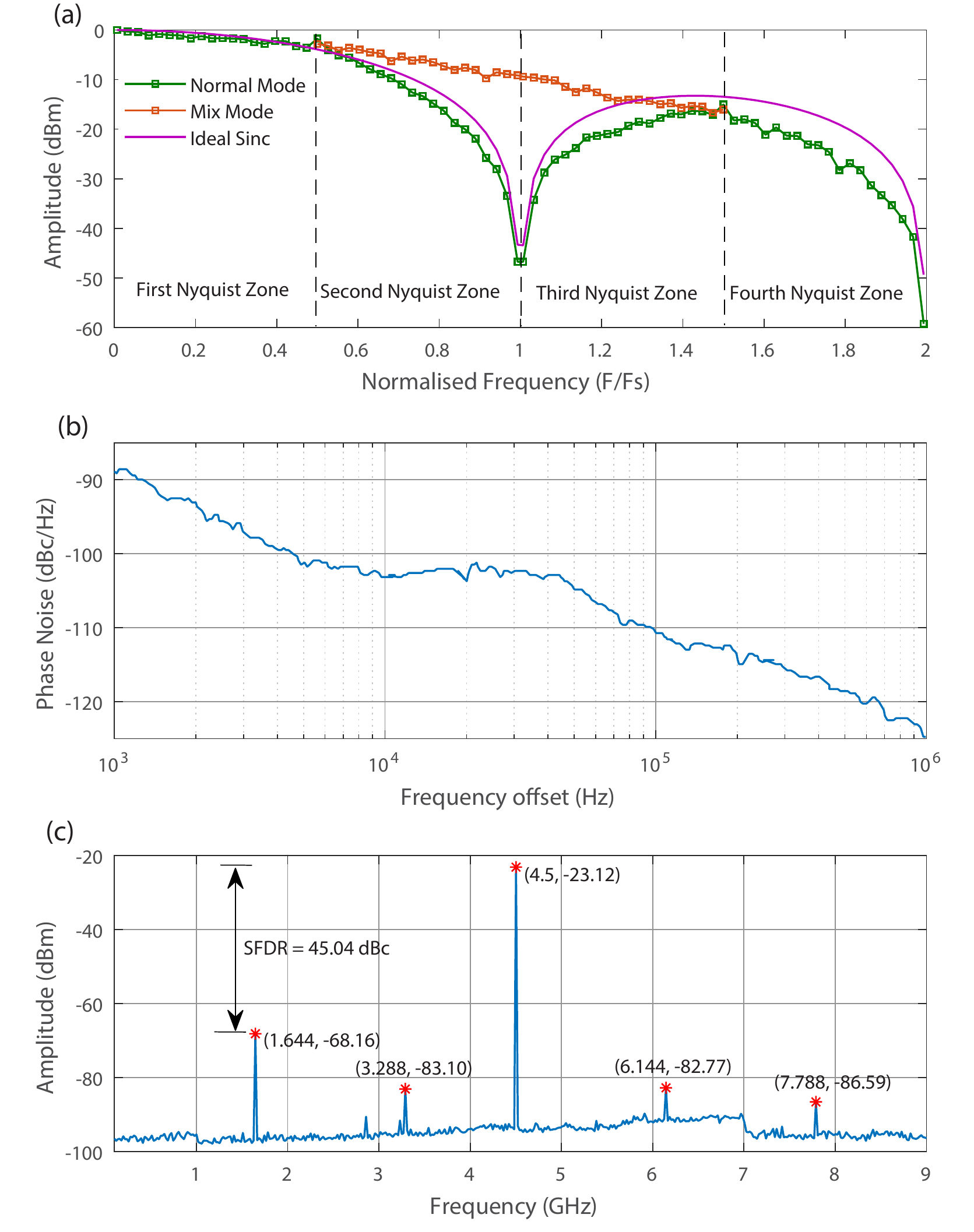}
\caption{RF performance evaluation of XCZU28DR:
(a) amplitude response of the DAC across various Nyquist zones in normal mode (NRZ) and
mix mode (RTC mode). A plot of sinc function in included for comparison. 
(b) Single sideband (SSB) phase poise of XCZU28DR at carrier frequency of 4.5 GHz. 
(c) A large span spectrum for the measurement of SFDR at 4.5~GHz carrier tone.}
\label{RFPerformance Fig}
\end{figure}

We use a signal analyzer (Rohde and Schwartz FSV-13) to 
perform the phase noise measurements. Figure~\ref{RFPerformance Fig}(b) 
shows the measurement of SSB phase noise plotted 
for different offset frequencies around the carrier 
frequency of $4.5$~GHz generated from the XCZU28DR device. 
The measured phase noise of $-102$~dBc/Hz at $4.5$~GHz 
carrier frequency at offset 10~kHz is comparable to 
the phase noise performance of standard test and 
measurement RF Signal generators \cite{sg394,tsg4100a,vsg60A,n5166b,shfqa}. 

While generation of signal using multi-zone Nyquist 
technique expands the scope of frequency domain capabilities 
of a DAC, the images generated in other zones need to 
be carefully suppressed to achieve a practically 
useful spurious-free dynamic range (SFDR) \cite{SFDR_sdr}. 
Figure~\ref{RFPerformance Fig}(c) shows the various spurs 
for a carrier tone of 4.5 GHz while using a {standard 
bandpass-filter minicircuit VBFZ-3590-S+(15542).} With this 
general purpose filter, we achieve a SFDR of nearly 45~dB. 
Our focus in this study has been on the generation of 
control signals for the superconducting qubits in the 
frequency range of 4-4.5~GHz. 

It is important to mention here that this value is currently limited 
by the choice of the filter, and can be further improved by using
tunable cavity or switchable filter banks \cite{borja_2_2010,tunablebpf}.

\begin{figure}[!htb]
\begin{center}
\includegraphics[scale=0.35]{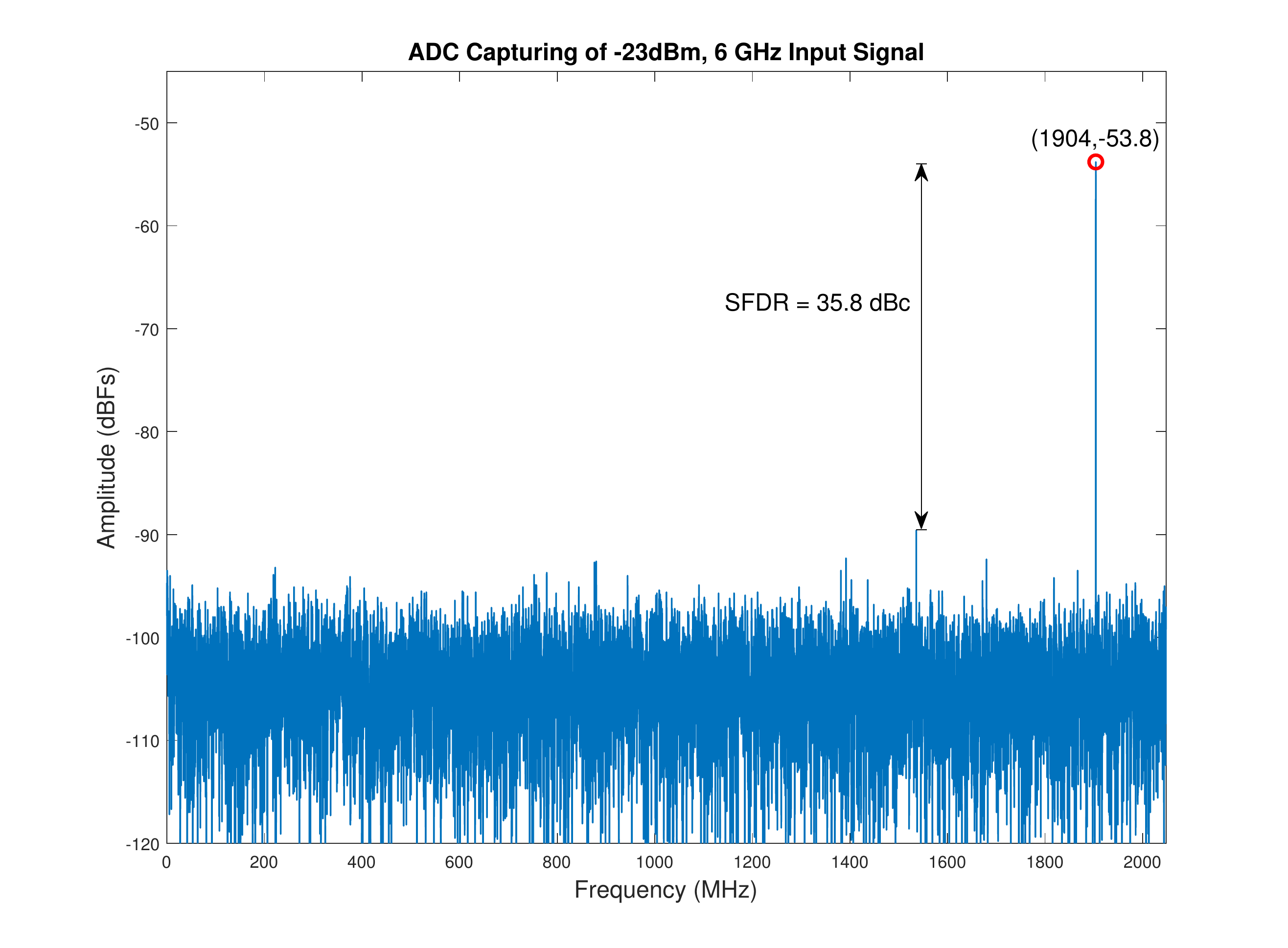}
\caption{Spectrum of 6 GHz signal acquired by ADC in mix mode  of operation}
\label{6GCapture}
\end{center}
\end{figure}

The ADCs are designed to directly measure the 
signals in higher Nyquist zones. The zone of 
operation is indicated to the digital calibration 
engine of RFSoC to ensure optimal performance 
of the ADC. A 6 GHz signal at -23~dBm is fed to 
ADC in mix-mode operating at a sampling rate 
of 4.096 GSPS. The resultant image would appear 
at $6000-4096 = 1904$~MHz. The frequency response 
of the signal captured by ADC is shown in 
Figure~\ref{6GCapture}.

\begin{figure}[!htb]
\begin{center}
\includegraphics[scale=0.36]{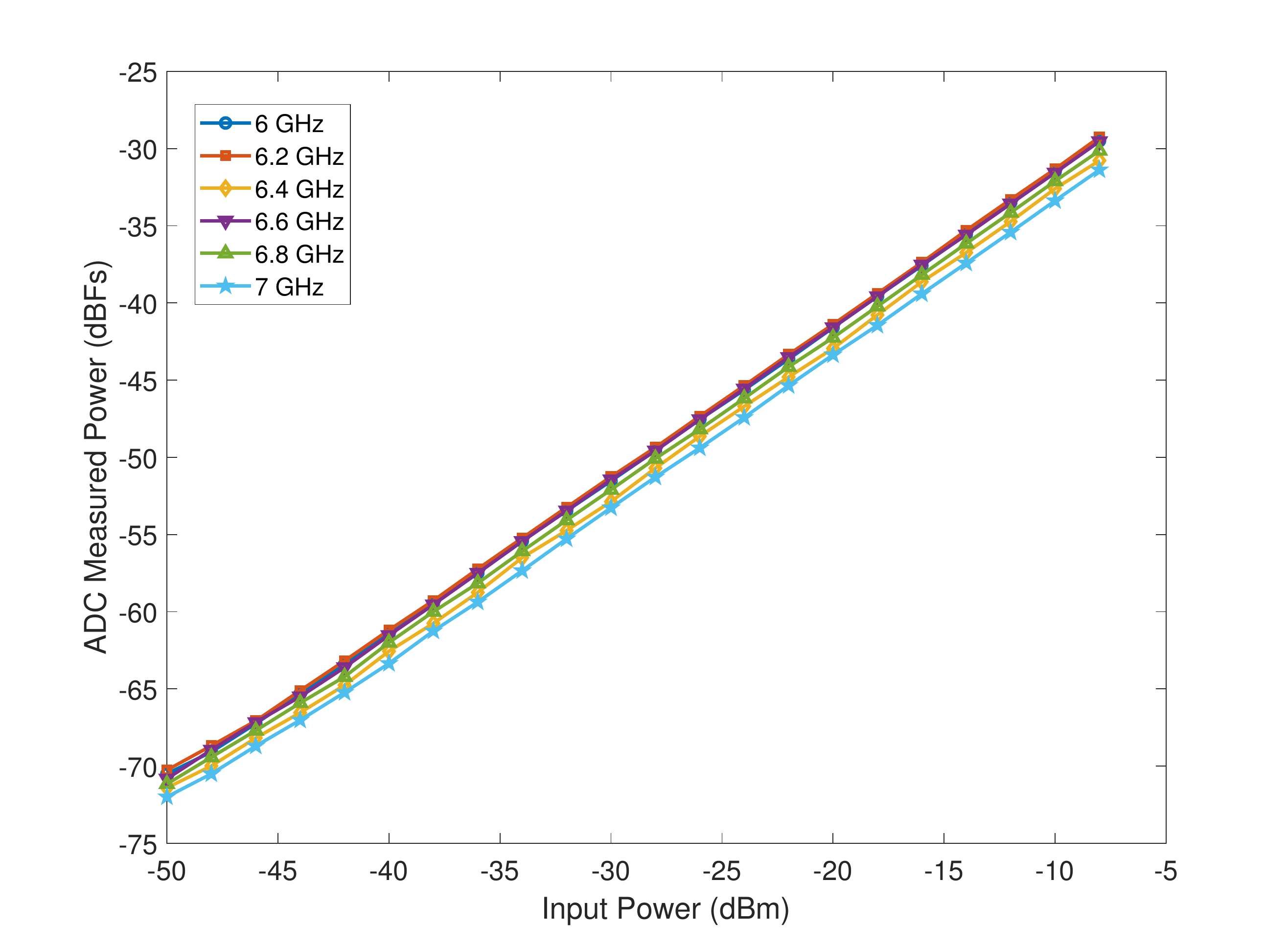}
\caption{Linearity of ADC from 6 GHz to 7 GHz}
\label{ADC_Linearity}
\end{center}
\end{figure}

To measure the linearity of ADC response in the Mix-mode, 
a signal is fed with varying input power for 
frequencies between $6-7$~GHz and the resultant 
amplitude (dBFs) is recorded. The measured amplitude at
the ADC shows a good linear behaviour with respect to the input power
as shown in Figure~\ref{ADC_Linearity}.

\begin{table*}[!htb]
\centering
\caption{Comparison of SQ-CARS with various state-of-the-art Quantum Control System platforms}
\resizebox{18.1cm}{!}{
\begin{tabular}{|ccc|c|c|c|c|c|c|}
\hline
\multicolumn{3}{|c|}{}                                                                                                                                                                                                        & Qubic\cite{xu2021qubic}  & QICK\cite{stefanazzi2022qick}                                                                   & Yang et.al\cite{yang2022fpga}                                                         & ICARUS-Q\cite{park2022icarus}  & Presto\cite{tholen2022measurement}                                                                & SQ-CARS                                                               \\ \hline
\multicolumn{3}{|c|}{Platform}                                                                                                                                                                                                & VC707  & ZCU111                                                                 & Kintex 7                                                           & HTG-ZRF16 & \begin{tabular}[c]{@{}c@{}}ZCU208/\\ ZCU216\end{tabular}              & ZCU111                                                                \\ \hline
\multicolumn{1}{|c|}{\multirow{7}{*}{Features}} & \multicolumn{1}{c|}{\multirow{2}{*}{Sampling Rate (GSPS)}}                                                        & DAC                                                     & 1.25   & 6.144                                                                  & 2                                                                  & 6.144     & 10                                                                    & 6.144                                                                 \\ \cline{3-9} 
\multicolumn{1}{|c|}{}                          & \multicolumn{1}{c|}{}                                                                                             & ADC                                                     & 1      & 4.096                                                                  & 1                                                                  & 1.96608   & 5                                                                     & 4.096                                                                 \\ \cline{2-9} 
\multicolumn{1}{|c|}{}                          & \multicolumn{2}{c|}{Multi Channel Sync}                                                                                                                                     & No     & \begin{tabular}[c]{@{}c@{}}using state \\ machine, No MTS\end{tabular} & \begin{tabular}[c]{@{}c@{}}using clock\\ distribution\end{tabular} & using MTS & using MTS                                                             & using MTS                                                             \\ \cline{2-9} 
\multicolumn{1}{|c|}{}                          & \multicolumn{2}{c|}{User Interface}                                                                                                                                         & Python & Python                                                                 & NA                                                                 & Python    & Python                                                                & Python                                                                \\ \cline{2-9} 
\multicolumn{1}{|c|}{}                          & \multicolumn{1}{c|}{\multirow{2}{*}{\begin{tabular}[c]{@{}c@{}}Microwave Synthesis \& \\ Capturing\end{tabular}}} & Mix-Mode                                                & No     & No                                                                     & No                                                                 & Yes       & Yes                                                                   & Yes                                                                   \\ \cline{3-9} 
\multicolumn{1}{|c|}{}                          & \multicolumn{1}{c|}{}                                                                                             & \begin{tabular}[c]{@{}c@{}}On-board\\ NCOs\end{tabular} & No     & No                                                                     & NA                                                                 & No        & Yes                                                                   & Yes                                                                   \\ \cline{2-9} 
\multicolumn{1}{|c|}{}                          & \multicolumn{2}{c|}{Information Processing}                                                                                                                                 & No     & Filtering                                                              & No                                                                 & No        & \begin{tabular}[c]{@{}c@{}}Template Matching,\\ Feedback\end{tabular} & \begin{tabular}[c]{@{}c@{}}Tunable Filtering,\\ Rotation,\\ Averaging\end{tabular} \\ \hline
\end{tabular}
}
\label{CompTab}
\end{table*}

\subsection{\centering Comparison with State-of-the-Art}
The proposed architecture has been implemented using 
Vivado 2020.2 and the configuration parameters of the 
experiment like the name of the experiment, pulse duration, 
time between pulses, etc., are being passed using SQ-CARS 
Python framework running on PYNQ v2.7. Table \ref{CompTab} 
show a comparison of the proposed SQ-CARS with the state-of-the-art Quantum control platforms. 

The major challenges in the design of an integrated control and readout system are the scalability, direct synthesis and capture of microwave signals, on-board data processing and an unified user interface. The existing approaches, such as  QubiC \cite{xu2021qubic}, QICK \cite{stefanazzi2022qick}, and Yang et al.\cite{yang2022fpga}, do not provide a comprehensive end-to-end solution for addressing all the challenges aforementioned. These approaches rely on discrete components and custom clock distribution modules at the baseband level, resulting in increased complexity and challenges in unified control. QICK does not make use of the on-board NCOs for generation and capture of microwave frequencies. These works use external analog mixers to up-convert the baseband signals. Their additional utilization of external analog mixers for RF conversion introduces problems such as leakage, periodic calibration etc. Although QICK achieves inter-channel phase synchronization using a custom state machine, it fails to address the variable delay caused by the RF-SoC pipeline due to NCOs and FIFOs and does not support MTS. ICARUS-Q\cite{park2022icarus} performs MTS but again like QICK, it does not utilize the on-board NCOs. This limits on-the-fly synchronized control of frequency and phase across different channels and leaves the variable delay caused by the RFSoC pipeline unaddressed.    

To overcome the limitations of the previous works, our work, SQ-CARS combines MTS with on-board NCOs to directly synthesize, control and capture phase-synchronized signals, demonstrating inter-channel synchronization among all data converters. We also developed a Python-based framework, by creating an abstraction layer that conceals the complexities of hardware implementation. It offers a user-friendly configuration interface to the physics user community, simplifying their interaction with the underlying hardware. Leveraging the existing RFdc Python code, we enhanced it by incorporating Multi-Tile Synchronization (MTS) functionality which enables on-the-fly synchronized control of frequency and phase of different channels. In addition, we developed Python classes for other hardware IPs within the programmable logic (PL), enabling seamless communication between the framework and hardware design modules. This streamlined approach enhances flexibility and ease of use for researchers and engineers working with the framework. To enable low-latency active feedback, on-board processing of acquired readout signals is essential. SQ-CARS incorporates on-board tunable filtering, rotation, and averaging blocks, which are absent in ICARUS-Q, Yang et al., and QubiC.

\begin{table*}[!htb]
\centering
\caption{Resource Utilization Comparison}
\resizebox{18.1cm}{!}{
\begin{tabular}{|c|c|cc|cc|cc|cc|c|}
\hline
\multirow{2}{*}{} & \multirow{2}{*}{Device} & \multicolumn{2}{c|}{Slice LUTs}                                                             & \multicolumn{2}{c|}{Flip Flops}                                                             & \multicolumn{2}{c|}{BRAMs}                                                               & \multicolumn{2}{c|}{DSPs}                                                               & \multirow{2}{*}{\begin{tabular}[c]{@{}c@{}}Information\\ Processing\end{tabular}} \\ \cline{3-10}
                  &                         & \multicolumn{1}{c|}{Available} & Consumed                                                   & \multicolumn{1}{c|}{Available} & Consumed                                                   & \multicolumn{1}{c|}{Available} & Consumed                                                & \multicolumn{1}{c|}{Available} & Consumed                                               &                                                                                   \\ \hline
Qubic \cite{xu2021qubic}            & XC7VX485T               & \multicolumn{1}{c|}{303600}    & \begin{tabular}[c]{@{}c@{}}Not\\ Reported\end{tabular}     & \multicolumn{1}{c|}{607200}    & \begin{tabular}[c]{@{}c@{}}Not\\ Reported\end{tabular}     & \multicolumn{1}{c|}{1030}      & \begin{tabular}[c]{@{}c@{}}Not\\ Reported\end{tabular}  & \multicolumn{1}{c|}{2800}      & \begin{tabular}[c]{@{}c@{}}Not\\ Reported\end{tabular} & No                                                                                \\ \hline
QICK \cite{stefanazzi2022qick}             & XCZU28DR                & \multicolumn{1}{c|}{425280}    & \begin{tabular}[c]{@{}c@{}}Not\\ Reported\end{tabular}     & \multicolumn{1}{c|}{850560}    & \begin{tabular}[c]{@{}c@{}}Not\\ Reported\end{tabular}     & \multicolumn{1}{c|}{1080}      & \begin{tabular}[c]{@{}c@{}}Not\\ Reported\end{tabular}  & \multicolumn{1}{c|}{4272}      & \begin{tabular}[c]{@{}c@{}}Not\\ Reported\end{tabular} & Filtering                                                                         \\ \hline
Yang et.al. \cite{yang2022fpga}      & XCKU060                 & \multicolumn{1}{c|}{331680}    & \begin{tabular}[c]{@{}c@{}}Not\\ Reported\end{tabular}     & \multicolumn{1}{c|}{663360}    & \begin{tabular}[c]{@{}c@{}}Not\\ Reported\end{tabular}     & \multicolumn{1}{c|}{1080}      & \begin{tabular}[c]{@{}c@{}}Not\\ Reported\end{tabular}  & \multicolumn{1}{c|}{2760}      & \begin{tabular}[c]{@{}c@{}}Not\\ Reported\end{tabular} & No                                                                                \\ \hline
ICARUS-Q  \cite{park2022icarus}        & XCZU29DR                & \multicolumn{1}{c|}{425280}    & \begin{tabular}[c]{@{}c@{}}212640\\ (50\%)\end{tabular}    & \multicolumn{1}{c|}{850560}    & \begin{tabular}[c]{@{}c@{}}Not\\ Reported\end{tabular}     & \multicolumn{1}{c|}{1080}      & \begin{tabular}[c]{@{}c@{}}810\\ (75\%)\end{tabular}    & \multicolumn{1}{c|}{4272}      & \begin{tabular}[c]{@{}c@{}}6\\ (0.14\%)\end{tabular}   & No                                                                                \\ \hline
Presto \cite{tholen2022measurement}           & XCZU48DR                & \multicolumn{1}{c|}{425280}    & \begin{tabular}[c]{@{}c@{}}Not\\ Reported\end{tabular}     & \multicolumn{1}{c|}{850560}    & \begin{tabular}[c]{@{}c@{}}Not\\ Reported\end{tabular}     & \multicolumn{1}{c|}{1080}      & \begin{tabular}[c]{@{}c@{}}Not\\ Reported\end{tabular}  & \multicolumn{1}{c|}{4272}      & \begin{tabular}[c]{@{}c@{}}Not\\ Reported\end{tabular} & \begin{tabular}[c]{@{}c@{}}Template Matching,\\ Feedback\end{tabular}             \\ \hline
SQ-CARS           & XCZU28DR                & \multicolumn{1}{c|}{425280}    & \begin{tabular}[c]{@{}c@{}}197271\\ (46.38\%)\end{tabular} & \multicolumn{1}{c|}{850560}    & \begin{tabular}[c]{@{}c@{}}194580\\ (22.87\%)\end{tabular} & \multicolumn{1}{c|}{1080}      & \begin{tabular}[c]{@{}c@{}}464\\ (42.96\%)\end{tabular} & \multicolumn{1}{c|}{4272}      & \begin{tabular}[c]{@{}c@{}}2244\\ (52.52)\end{tabular} & \begin{tabular}[c]{@{}c@{}}Filtering,\\ Rotation, Averaging\end{tabular}          \\ \hline
\end{tabular}
}
\label{Utilization}
\end{table*}

In Table \ref{Utilization}, we compare the resource utilization of SQ-CARS with existing platforms. Except ICARUS-Q, no other work has reported resource utilization. Hence, it is written `Not Reported' indicating the absence of the report of the corresponding parameter in the literature. SQ-CARS achieves waveform playtime of 80us, a crucial experimental parameter, by leveraging on-board NCOs. This is significantly higher than ICARUS-Q, which only offers a playtime of 10us while utilizing similar  BRAM resources per channel. Another main contribution lies in the on-board information processing pipeline employed by SQ-CARS, which is absent in ICARUS-Q. SQ-CARS consumes 46.38\% of available LUTs and 42.96\% of available BRAM resources which leaves ample room for expanding the system's capacity to accommodate a larger number of qubits, ensuring scalability.

The Power consumption of SQ-CARS has been generated using Vivado 2020.2 enabling the power optimisation feature. SQ-CARS consumes consumes 1.456 W of Static power and 20.353 W of Dynamic power, out of which only 15\% is due to Logic. The rest of the power is majorly due to hard IPs like RF-DACs, RF-ADCs etc. These power numbers are unreported by any of the previous works. 

In conclusion, SQ-CARS offers a comprehensive end-to-end solution, with the set of features including direct microwave synthesis and capture, long-term inter-channel phase synchronization, a scalable and flexible architecture, a digital pipeline for information processing, and a user-friendly unified interface, all on a single platform. These features, except for those found in Presto \cite{tholen2022measurement}, enable SQ-CARS stand apart from the previous approaches.

\begin{figure*}
\centering
\includegraphics[width = 100mm]{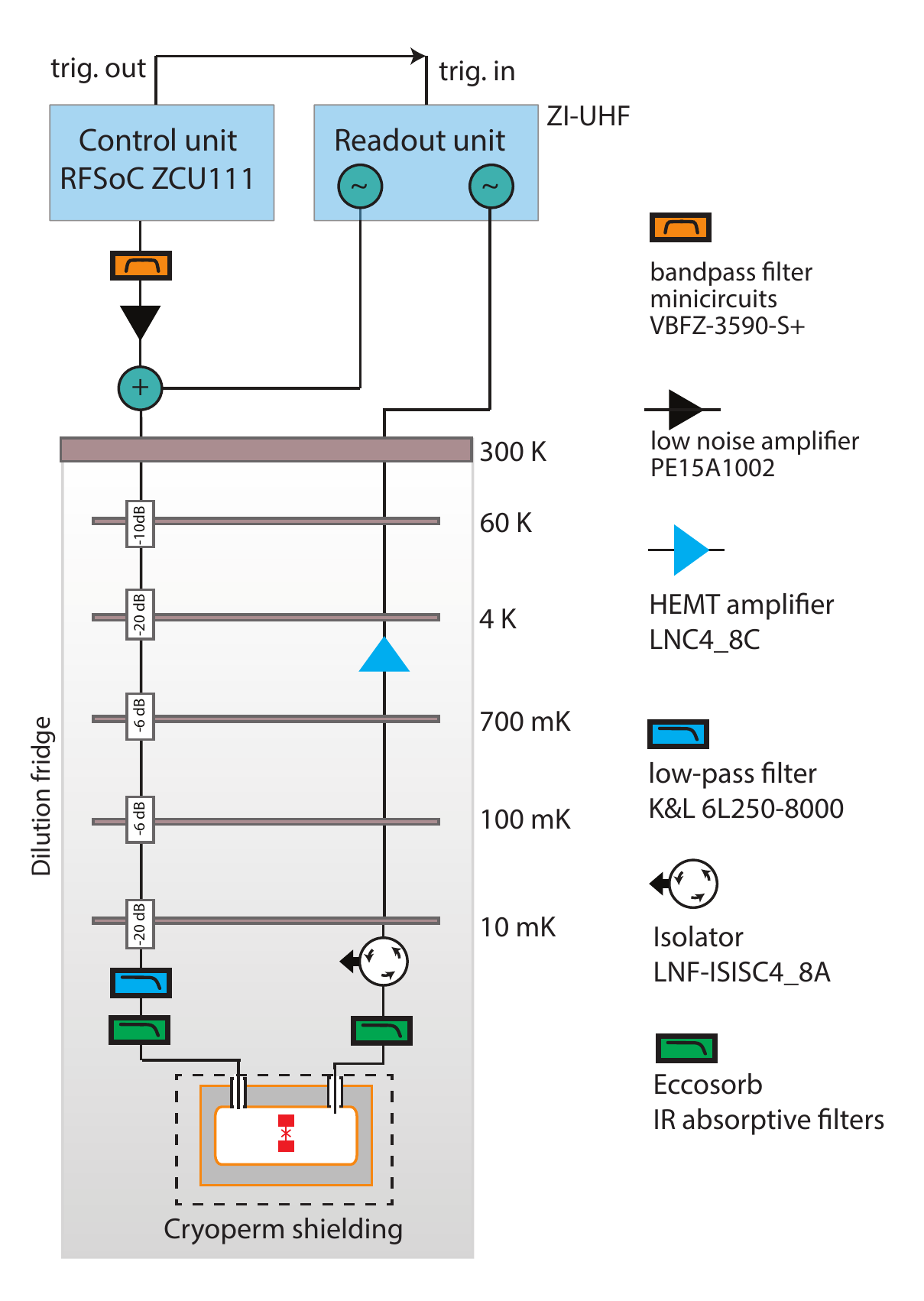}
\caption{Experimental measurement setup of superconducting qubit devices : 
The input lines are used for qubit state manipulation and to 
send cavity probe signals. First, the control signal was boosted up 
by using a Pasternack low-noise amplifier, and then it was added 
together with the readout signal. Both the signals then traveled 
to the device through the series of attenuators, tubular low pass 
filter, and homemade IR filter. The output signal from the cavity 
is then amplified by a HEMT amplifier mounted on the 4K stage 
and then the cavity quadratures were measured directly using the
ZI-UHF lock-in amplifier. }
\label{setup}
\end{figure*}
\section{Coherence measurements of a superconducting qubit}

To benchmark the performance of our technique for the 
direct generation of control signal pulses, we use a superconducting 
transmon qubit coupled to a 3D waveguide copper cavity. 
A fixed-frequency transmon \cite{paik_observation_2011} 
qubit was fabricated using standard lithography processes 
on an intrinsic silicon wafer. The whole qubit-cavity setup 
was mounted to a dilution refrigerator's base flange (10~mK) with 
various attenuators and filters on the input microwave lines. 
To benchmark the performance of  RFSoC, we test its performance 
with two different transmon qubit devices named D1 and D2. 
A schematic of the full measurement setup is shown in Fig.~\ref{setup}.
By carrying out cavity-qubit spectroscopy, we find the 
cavity frequency $\omega_c/2\pi=5.995$~GHz, frequency of 
qubits D1 and D2 are approximately 4.2 and 4.7 GHz, respectively. 
We then characterize the performance of the RFSoC by 
generating the microwave pulses to carry out time domain 
characterization of qubits. To demonstrate the applicability 
of the directly synthesized control pulses, we carry out 
time-domain coherence measurements of the
superconducting qubit. Generally, the time-domain experiments 
are done in two steps. First the calibration of 
microwave pulses is carried out. It may then followed by some
basic characterization experiment.

\subsection*{\centering A. Readout and control}

To generate the microwave control pulses for the qubit, 
we use RFSoC DAC. As the qubit frequencies fall in the 
second-Nyquist zone, we use a wide-band Pasternack (PE 15A1002) 
amplifier to boost the signal 
amplitude to overcome the attenuation before sending it 
into the fridge.
The baseband readout signal was generated using UHF lock-in 
amplifier from Zurich Instruments with AWG option.
We used a home-built frequency up-converter and down-converter 
setup to generate the microwave readout pulse and to demodulate 
the readout signal coming from the fridge. 
We then determine the in-phase and quadrature components 
of the readout pulse and the subsequent qubit state determination. 
To address the issue of cross-platform triggering, we generated 
a trigger from ZCU111 analog output channel and used it to arm 
the sequencer on UHF-AWG. This straightforward handshaking 
allows us to correct any trigger latency by advancing the 
trigger generation on ZCU111. Moreover, it allows defining 
simple ``loops" on ZCU111 so that the entire experiment can be 
controlled from the board. It is important to mention 
that, in principle, the fast ADC channels available on 
ZCU111 can be used for qubit readout and the entire 
control and readout part can be done on the same board.

\subsection*{\centering B. Rabi spectroscopy}

\begin{figure*}
\centering
\includegraphics[width = 140mm]{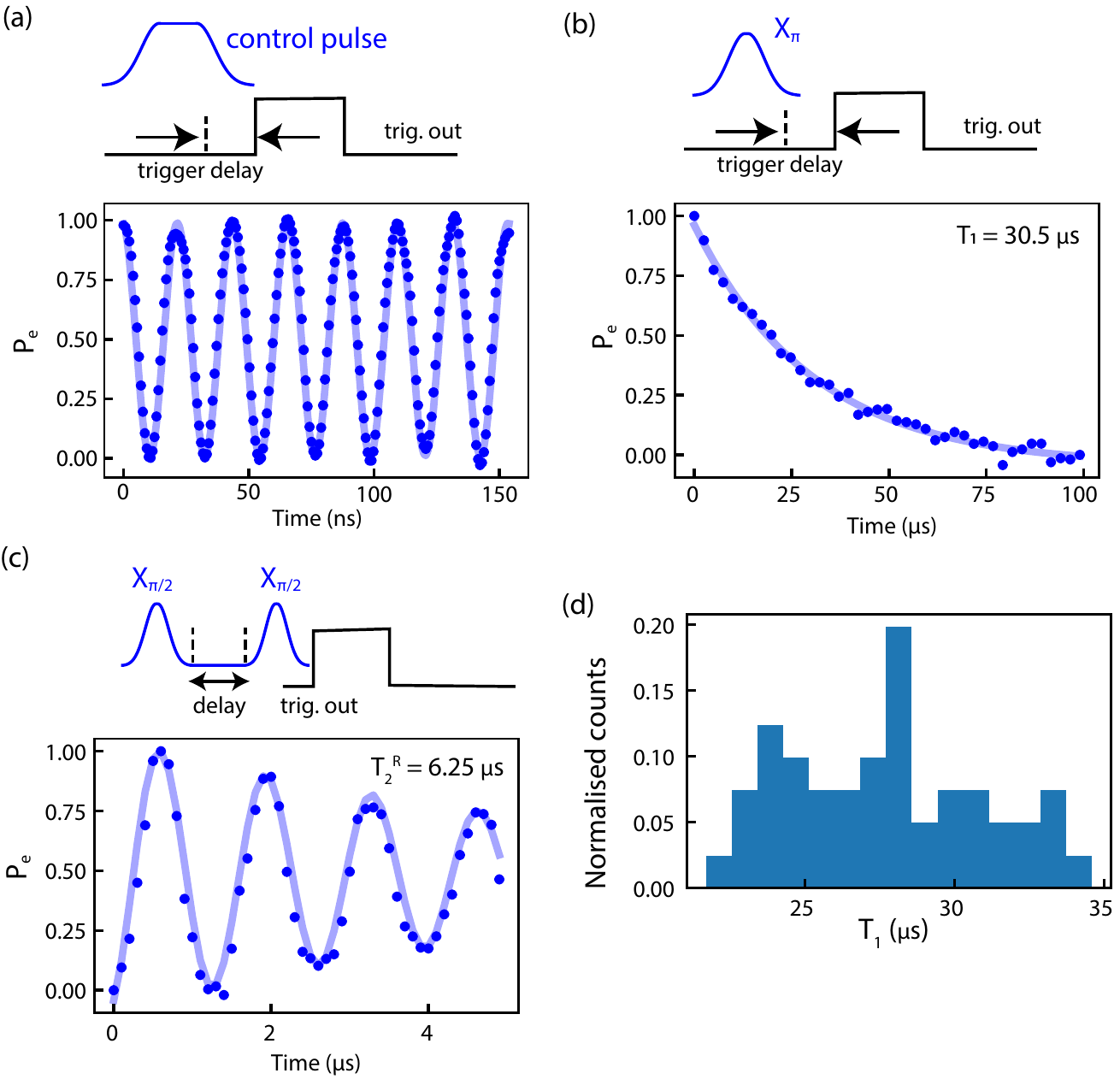}
\caption{Time-domain control of a transmon qubit (D1) 
using ZCU111. Each measured data point in these panels is a result from 50,000
shots. (a) Rabi oscillations 
between the qubit’s ground $\ket{0}$ and first excited state $\ket{1}$ as 
a the duration of control pulse is varied. (b) Measurement of the energy 
relaxation time $T_1$. The blue line indicates an exponential fit 
yielding $T_1=30.5~\mu$s.
(c) Measurement of Ramsey dephasing time ($T_2^R$). The top part of the panel 
shows the pulse sequence. From the fit indicated by solid line, we extract 
Ramsey dephasing time $T_2^R=6.25~\mu$s and detuning $\Delta = 0.8$~MHz. 
(d) Histogram showing the statistical variation between 47 $T_1$ measurements 
taken over a period of 24 hours.}
\label{T1T2}
\end{figure*}

It involves measuring the excited state qubit population while varying 
the amplitude/length of the Gaussian pulse resonant with the qubit frequency, 
the measurement colloquially known as power/time Rabi measurement. 
We implemented this experiment on device D1, by inserting a rectangular 
pulse with a variable number of samples at the center of the $X_{\pi}$ 
Gaussian pulse. We use 260~ns long Gaussian pulses for 
these measurements with a standard deviation of 65~ns. 
The duration of the qubit control pulse can be adjusted 
by adjusting the number of sample points in the rectangular 
pulse. The top part of Fig.~\ref{T1T2}(a) panel shows 
the pulse sequence. Any trigger latency between ZCU11 
and the readout setup (UHF) is adjusted by adjusting 
the trigger delay. The bottom panel shows the coherent 
time oscillations as the duration of the rectangular portion 
of the control pulse is increased.
The measurement of Rabi oscillations in time and power domain allows
to calibrate the microwave pulses necessary for qubit operation.

\subsection*{\centering C. Relaxation time measurement }

After getting the calibrated $X_{\pi}$ from the 
above measurement, it is straightforward to carry out 
energy relaxation time measurement. The setup is same as the 
Rabi oscillation measurement setup with a minor change in the 
pulse sequence. Here we first send a calibrated  $X_{\pi}$ pulse 
and then wait for time $t$, before performing a readout as 
shown in Fig.~\ref{T1T2}(b). Due to on-the-fly configuration 
of the delay time, the entire measurement can be controlled 
by ZCU111 unit. For a given wait time, we generate 
50,000 $X_{\pi}$ pulses at a repetition rate of 5~kHz 
and determine the excited state population. Fig.~\ref{T1T2}(c)
shows the result from $T_1$ measurements of D1 device. 
The value of $T_1$ was extracted by fitting the following 
equation $A + Be^{-t/T_1}$.
The results from the $T_1$ experiment along with the fits, 
yielded a $T_1$ of approximately 30.5 $\mu$s.

\begin{figure*}
\centering
\includegraphics[width = 140mm]{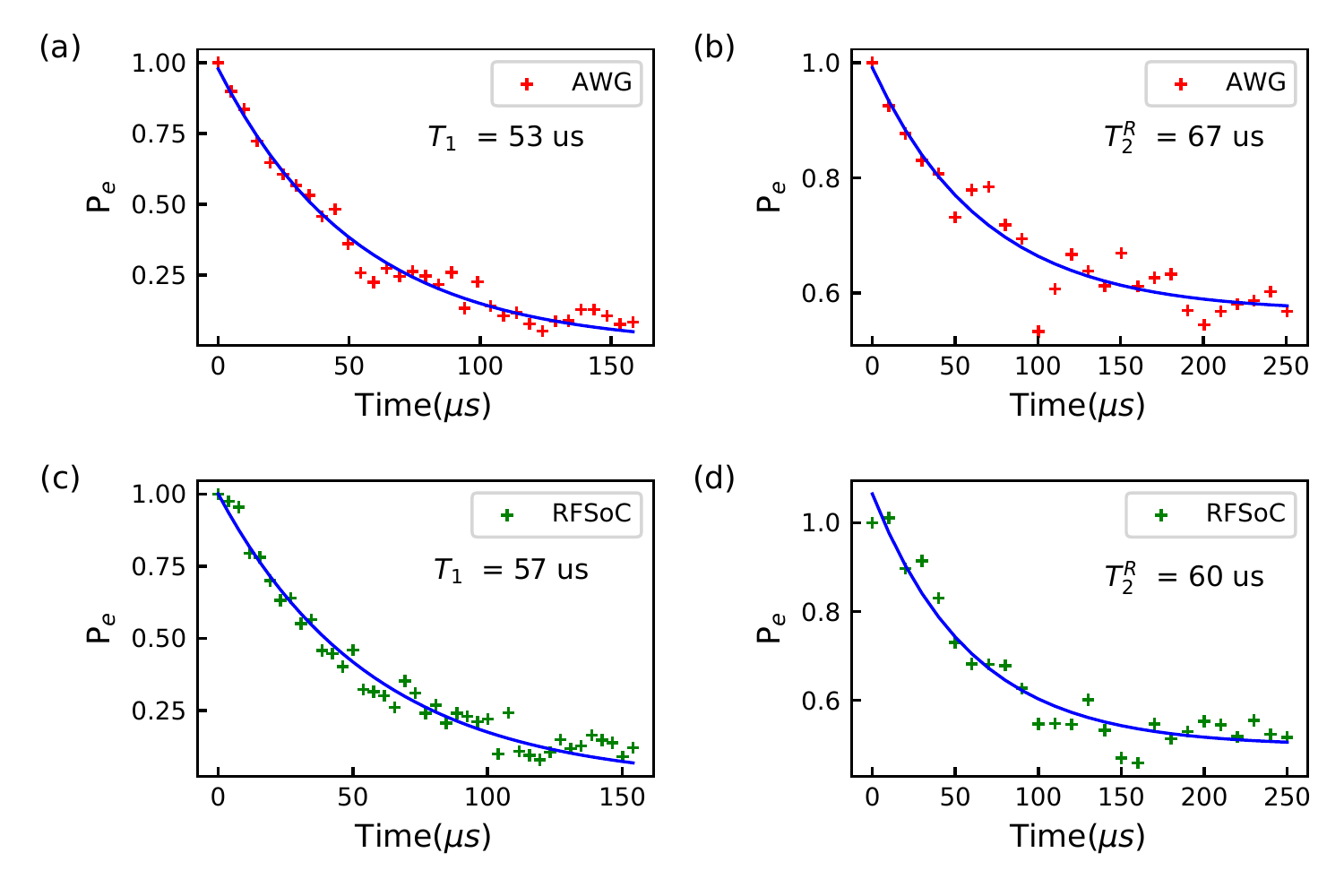}
\caption{Comparison of the measurements of $T_1$ and $T_2^R$ from a transmon qubit (D2) 
using traditional AWG and RFSoC.
Panel (a, b) show $T_1$ and $T_2^R$ measurement obtained from traditional AWG. 
The solid line indicates an exponential fit 
yielding $T_1=53~\mu$s and $T_2^R=67~\mu$s.
Panel (c, d) show the measurement of the $T_1$ and $T_2^R$ using RFSoC DAC. 
The solid line indicates an exponential fit 
yielding $T_1=57~\mu$s and $T_2^R=60~\mu$s. 
For Ramsey measurements, we drive the qubit on resonance and thus 
it does not show any beating oscillations. 
Each measured data point in these panels is a result from 50,000
shots.}
\label{T1T2_re}
\end{figure*}

\subsection*{\centering D. Ramsey spectroscopy }
Next, we carry out the Ramsey experiment, a measurement 
of dephasing rate. The protocol requires the generation of 
two $X_{\pi/2}$ pulses with a variable time-delay between them, 
and it is followed by the qubit measurement. 
A schematic of the pulse sequence and result from the 
Ramsey fringe experiment is shown in Fig.~\ref{T1T2}(c).
As the experiment can be completely controlled by the ZCU111, 
the procedure remains similar to the $T_1$ measurements. 
Here the trigger for the readout unit is aligned with the end
of the second $X_{\pi/2}$ pulse and a delay between the 
two $X_{\pi/2}$ pulses is varied. By fitting the measured 
curve of the Ramsey fringes using  $A + B \cos(2\pi\Delta t+\phi) e^{-t/T_2^R}$, 
one can obtain the values for $T_2^R$ and detuning $\Delta$, respectively. 
In this context, the detuning represents the deviation of the microwave pulse 
frequency from the qubit frequency. 
For the device D1, we measured $T_2^R$ of 6.25 $\mu$s and detuning 
of 0.8 MHz. It is important to point out here that since 
both the $X_{\pi/2}$ pulses are generated by the internal 
digital IQ-mixer and common NCO, a well-defined phase 
relation holds between the two pulses. Such functionality makes 
it very easy to implement where \textit{X-} and \textit{Y-} qubit rotations
can be made by controlling the signals on \textit{I} and \textit{Q} inputs 
of the digital mixer.

\subsection*{\centering E. Standard AWG vs RFSoC }

To compare the performance of RFSoC and standard
test and measurement instruments, we compare the
results of qubit characterization, $T_1$ and $T_2^R$, on 
the same device during the same cooldown run.
The conventional measurements were carried out by generating 
the drive pulses using the single-sideband modulation
technique and a vector signal generator. The baseband drive signals
were generated using a commercially available AWG. 
In the Fig.~\ref{T1T2_re}, we show the comparative results 
from another device (D2) by measuring it using general test and 
measurement equipment and RFSOC. 
We do not observe any statistically significant difference 
between the results. Given the ease of the measurements, as no periodic 
calibration for sideband suppression is necessary, we also repeated 
the $T_1$ measurements of device D1 over 24 hours and recorded 
statistical variations in the $T_1$ measurements.
Fig.~\ref{T1T2}~(d) shows a histogram of 47 $T_1$ measurements 
from the same device.

\section{Conclusion}

In this work, we have showcased the performance and capabilities of our integrated framework SQ-CARS, designed to cater to the demanding requirements of superconducting quantum systems. Our framework not only allows for direct generation and capture of microwave signals, but also supports real-time information processing, providing capability for active feedback within the system. What sets our framework apart is its scalability, configurability, and user-friendly nature, all achieved while keeping costs low. By providing an accessible room temperature control system, encompassing the capabilities and features detailed in this paper which is freely available on GitHub, invites further collaboration and research in the field. Furthermore, the versatility of our techniques allows for easy expansion to accommodate a larger number of channels, opening up possibilities for experimentation and research. While our primary focus has been on the superconducting quantum devices community, the implications of our work extend beyond this domain. Researchers working with quantum computing using semiconductors, Nitrogen Vacancy (NV) centers, and trapped-ion systems will also find value and relevance in our platform. 

Looking ahead, our future work will focus on further enhancements to the proposed system. To reduce overall duty cycle, we would be incorporating on-board qubit state detection and active feedback mechanisms to send corrective pulses without the intervention of the user. We will also work on implementing the capability of direct play of longer waveforms from faster DDRs, faster streaming of data by integrating Ethernet to PL. 

Unlike classical processors, quantum circuits face a substantial input-output bottleneck. Each qubit in a quantum computer is individually governed by external circuitry, which introduces both noise and heat to the qubit system\cite{krinner2019engineering}. Currently, the control of superconducting qubits relies on the application of pulsed microwave signals. The generation and routing of these control pulses involve considerable experimental overhead, encompassing both room-temperature and cryogenic electronics hardware. This includes, but is not confined to, coherent microwave waveform generators, amplifiers, as well as coaxial lines and signal conditioning elements required to transmit these signals into the low-temperature experimental environment. While the brute-force scaling of existing technology may suffice for moderately-sized superconducting qubit systems, the control of large-scale systems necessitates fundamentally novel approaches\cite{reilly2019challenges}. In this scenario, a noteworthy concept to consider is the integration of all crucial features onto a single chip, enabling the possibility of relocating the RFSoC (Radio Frequency System on a Chip) device from room temperature to the dilution refrigerator. This relocation would result in improved signal-to-noise ratio (SNR) and a reduction in the number of connections to room temperature electronics. Such a transition would offer an implementation of cryogenic control electronics for larger superconducting Qubit systems.  

\section{Acknowledgment}
Authors thank Baladitya Suri for valuable discussions. 
U.S. is supported by PMRF, Govt of India. C.S.T and V.S. acknowledge 
the support received under the Institute of Eminence (IoE)
scheme of Govt. of India. The authors acknowledge the support under the CoE-QT by
MEITY and QuEST program by DST, Govt. of India.
The authors acknowledge device fabrication 
facilities at CENSE, IISc, Bangalore, and central facilities at the Department 
of Physics funded by DST.

\end{document}